\documentclass[twocolumn,superscriptaddress,floatfix,preprintnumbers,nofootinbib,prd,10pt]{revtex4-2}
\usepackage{docs}%
\usepackage{bm}%
\usepackage[colorlinks=true,linkcolor=blue, citecolor=teal]{hyperref}
\pdfoutput=1
\usepackage{graphicx}
\usepackage{xcolor,colortbl}
\usepackage{subfig}
\usepackage{dcolumn}
\usepackage{mathtools}
\usepackage{bm}
\usepackage{amssymb}
\usepackage{amsmath}
\usepackage{epsfig}
\usepackage[toc,page]{appendix}
\usepackage{slashed}
\usepackage{hhline}
\usepackage{color}
\usepackage{amsmath} 
\usepackage{upgreek} 
\usepackage{bm} 
 \usepackage{multirow}
\usepackage{url}
\usepackage{empheq}
\def\be{\begin{equation}}
\def\ee{\end{equation}}
\def\bea{\begin{eqnarray}}
\def\eea{\end{eqnarray}}
\def\ba{\begin{array}}
\def\ea{\end{array}}
\def\ben{\begin{enumerate}}
\def\een{\end{enumerate}}

\def\bw{\begin{widetext}}
\def\ew{\end{widetext}}

\newcommand{\dsl}{\pa \kern-0.5em /} 
\def\a{\alpha}
\def\b{\beta}
\def\g{\gamma}

\def\l{\lambda}

\def\n{\nu}

\def\pa {\partial}

\begin{document}
\allowdisplaybreaks

\title{Pole-skipping and chaos in D3-D7 brane systems}

\author{Banashree Baishya}
\email[E-mail: ]{b.banashree@iitg.ac.in}
\affiliation{Department of Physics, Indian Institute of Technology Guwahati, Assam-781039, India}

\author{Sayan Chakrabarti}
\email[E-mail: ]{sayan.chakrabarti@iitg.ac.in}
\affiliation{Department of Physics, Indian Institute of Technology Guwahati, Assam-781039, India} 

\author{Debaprasad Maity} 
\email[E-mail: ]{debu@iitg.ac.in}
\affiliation{Department of Physics, Indian Institute of Technology Guwahati, Assam-781039, India} 

\author{Kuntal Nayek} 
\email[E-mail: ]{nayek.kuntal@gmail.com}
\affiliation{Department of Physics, Indian Institute of Technology Guwahati, Assam-781039, India}

\begin{abstract}
    In this paper, we analyse the pole-skipping phenomena of finite temperature Yang-Mills theory with quark flavors which is dual to D3-D7 brane systems in bulk. We also consider the external electric field in the boundary field theory which is dual to the world volume electric field on the D7 brane. We will work in the probe limit where the D7 branes do not back-react to the D3 brane background. In this scenario, we decode the characteristic parameters of the chaos namely, Lyapunov exponent $\lambda_{L}$ and butterfly velocity $v_b$ from the pole-skipping points by performing the near effective horizon analysis of the linearised Einstein equations. Unlike pure Yang-Mills, once charged quarks with a background electric field are added into the system, the characteristic parameters of the chaos show non-trivial dependence on the quark mass and external electric field. We have observed that $\lambda_{L}$ and $v_b$ decreases with increasing electric field. We further perform the pole-skipping analysis for the gauge invariant sound, shear, and tensor modes of the perturbation in the bulk and discuss their physical importance in the holographic context. 
\end{abstract}


\maketitle

\section{Introduction}
Theoretical understanding of chaos in the framework of quantum field theory has been the subject of interest for many years. The discovery of Gauge/Gravity duality \cite{Maldacena:1997re}, which provides an intriguing relation between the physics of black holes and quantum chaos, leads to a spate of research work on this field in the holographic framework. This duality provides a framework where a strongly coupled many-body quantum system living on the boundary of an AdS space can be understood from the properties of black holes located in the bulk of AdS. 
Generally, to measure quantum chaos, the \textit{out of time-ordered correlation} (OTOC) function has been proven to be an interesting field theoretic diagnostic tool. In the gravitational bulk, OTOC is calculated in an eternal black-hole geometry \cite{Maldacena:2001kr} when shock waves collide near the black hole horizon \cite{Shenker:2013pqa, Shenker:2013yza, Shenker:2014cwa, Roberts:2014isa}. In an interesting way, the same chaotic nature of a many-body system can be captured in the properties of a two-point energy density correlation function at the \textit{pole-skipping} \cite{Blake:2017ris, Blake:2018leo, Grozdanov:2018kkt, Ahn:2019rnq} point, and that has been shown to be holographically related to the near-horizon properties of the metric perturbation in bulk. Pole-skipping (P-S) occurs when lines of poles and zeroes of retarded Green's function intersect—a would-be pole gets skipped! At these specific values of the frequency and momentum (P-S point), Green's function is not uniquely defined \cite{Blake:2019otz, Natsuume:2019xcy}. This specific value of the frequency ($\omega_*$) and momentum ($k_*$) in the energy-density two-point function is connected to the measure of chaos \cite{Blake:2017ris, Blake:2018leo} through the following relations:
\begin{equation}\label{chaos}
\omega_{*}=i\lambda_{L},\hspace{2cm}k_{*}=\frac{i\lambda_{L}}{v_b} ,
\end{equation}
where $\lambda_{L}$ is the Lyapunov exponent and $v_b$ is the butterfly velocity. Over the years, a large number of studies have explored the chaotic nature of the holographic systems utilizing this P-S method in the black hole background \cite{Blake:2021hjj, Ahn:2019rnq, Wu:2019esr}, plasma physics \cite{Sil:2020jhr, Li:2019bgc, Amano:2022mlu}, conformal field theories \cite{Ramirez:2020qer, Haehl:2019eae, Das:2019tga}, topologically massive gravity \cite{Liu:2020yaf}, holographic system with chiral anomaly \cite{Abbasi:2019rhy}, little string theory \cite{Mahish:2022xjz}, with stringy corrections \cite{Grozdanov:2018kkt}, higher derivative corrections \cite{Natsuume:2019vcv, Wu:2019esr, BAISHYA2024116521}. This phenomenon has been extensively studied in various contexts and explored in different directions in \cite{Ceplak:2021efc, Yuan:2020fvv, Kim:2020url, Grozdanov:2020koi, Jeong:2021zhz, Abbasi:2020xli, Kim:2021xdz, Jeong:2022luo, Wang:2022mcq, Wang:2022xoc, Yuan:2023tft, Natsuume:2023lzy, Baishya:2023xbj, Yadav:2023hyg}. The QCD theory is very interesting to study quantum chaos \cite{qcdchaos, Akutagawa_2018, Shukla:2023pbp}. In the unquenched QCD, i.e., in the presence of light quarks, the chaotic nature of the system is expected to be different from the YM theory. Initially, the light quarks will be bounded in the form of neutral meson. But once these mesons melt into the charged quarks and anti-quarks, the Coulomb's interaction among these charged candidates is supposed to decrease the chaotic nature. This kind of effect in quantum chaos is very much new and interesting to study. On the other hand, if we apply an external electric field, the quark/anti-quark separation in meson increases and finally dissociates. This leads to a phase transition between the neutral meson-dominated insulating phase and the charged quark/anti-quark-dominated conducting phase. The chaotic parameters of the theory are also expected to show non-trivial change through this transition. In our present work, we represent this novel feature of quantum chaos.

In this paper, we study pole-skipping and the associated behaviour of the characteristic parameters of quantum chaos in 3+1-dimensional Yang-Mills theory in the presence of quark flavors at a finite temperature. The chaotic nature of the YM theory with quenched flavor is well explored. But, it is still very exciting to see the effect of flavor on the chaotic properties of the unquenched theory. In holography, the dual gravity theory of such a system can be achieved by probing D7 branes in the black-D3 brane background \cite{Karch:2002sh}. The Lagrangian of the probe brane vanishes at some special radial distance away from the D3 brane horizon, which will act as the effective horizon of the open-string induced metric on the D$7$ brane world volume. The emergence of such an effective horizon is expected to give rise to non-trivial effects, and indeed alter the chaos dynamics of the boundary theory in terms of characteristic parameters $(\lambda_{L}, v_b)$. 
 
Further, the probe branes can support the background Maxwell's field, and in the dual theory that will correspond to interacting charged flavors with the electric field. However, for the perturbative consideration, we must maintain the electric field in a very small range. This configuration is expected to add additional characteristic changes in the diagnostic parameters of chaos. Without the flavor D7 branes, a stack of black D3 branes in supergravity limit gives rise to an AdS-blackhole background in the bulk, corresponding to the Yang-Mills theory on the boundary. In the paper \cite{Shenker:2013pqa}, it was first shown that the AdS black holes are maximally chaotic, and hence the Lyapunov exponent of the boundary theory saturates the Maldacena-Shenker-Stanford (MSS) bound \cite{Maldacena:2015waa}. In the presence of Maxwell's field in the world volume of probe D7 brane, we found the dual theory to be no longer maximally chaotic, and the Lyapunov exponent depends on the electric field and the butterfly velocity depends on both the electric field and the number of flavors $N_f$ in the dual system. In this paper, we have explicitly shown how the electric field affects the Lyapunov exponent by pole-skipping analysis and this exponent obeys the MSS bound.

\par Non-trivial dependence of the chaotic parameters on the background electric field can be understood in terms of the behaviour of the meson states under the field. The light quark/anti-quark bound state in the boundary theory gives rise to the meson spectrum. The mesons are charged neutral in this case. But the constituent quark and anti-quark are positively and negatively charged respectively. Without any external electric field, the neutral meson states will not get affected in their probe limit. But, in the presence of an external electric field, the quark/anti-quark pair of mesons separates away by reducing its binding energy. Further, these charged quarks can have strong interaction with the background gluons in this situation. The motion of the neutral particle in the neutral background is more chaotic than the motion of the charged particle in the charged background. Because in the presence of the background charge, the charged particle encounters restoring drag due to Coulomb's interaction. Therefore the chaotic behaviour is expected to reduce due to the applied electric field. Further, the corresponding Lyapunov exponent is affected by the electric field. All these new physical effects can strongly influence the Yang-Mills dynamics. Lowering the binding energy causes the mesons to melt at a lower value of temperature, and at that point, the characteristic parameters of the chaos reduce. These new features influencing quantum chaos have been discussed in this present literature.

Apart from deriving the parameter of the chaos, we further calculate the P-S points associated with the hydrodynamic modes. With the choice of gauge invariant variables, we have constructed the master equations for sound, shear, and tensor channels perturbation variables \cite{Kodama:2003jz}. Expanding those equations near the effective horizon induced by the D7 brane, we have again calculated the P-S points in all three channels and discussed their relation with 
hydrodynamic transport and their dependence on the tuning parameters such as external electric field $E_0$, the number of flavors $N_f$ and quark mass $m_q$.

We have arranged the paper as follows: in section \ref{sec2}, we have given a brief discussion on the black D3 brane and constructed the embedding background. We have calculated the Lyapunov exponent and the butterfly velocity in section \ref{sec_PS}. In section \ref{sec_master}, we have calculated the pole-skipping points from the master equation in the perturbed channels. Finally, in the last section, \ref{conclusions}, we have concluded our work with some future directions.

   \section{Gravitational background}\label{sec2}
The type II superstring theory, at a low energy limit, reduces to the ten-dimensional supergravity action. The D$p$ branes are the solution to this action, where $p$ is the spatial dimension along the brane. Branes having the Arnowitt-Deser-Misner (ADM) mass equal to its charge are called extremal or Bogomol'nyi-Prasad-Sommerfield (BPS) branes. We consider non-extremal D$3$ brane in the bulk. Then, we construct a decoupled geometry where $N_c$ number of D$3$ branes are closely located. The decoupled geometry is the Schwarzschild-AdS$_5$ black-hole geometry. The corresponding boundary theory is $\mathcal{N}=4$ Yang-Mills theory at finite temperature with $N_c$ color charge. We further consider $N_f$ number of BPS D$7$ probe branes embedded in the bulk in the presence of Maxwell's field. According to the AdS/CFT dictionary, on the boundary we have the finite temperature $SU(N_c)$ $\mathcal{N}=4$ Yang-Mills theory with probe hypermultiplet quark flavors \cite{Karch:2002sh}. We will be working in the probe limit $N_f << N_c$ so that the probe brane does not back-react on the background geometry.

\subsection{Review of black D3 brane solution}
In this sub-section, we will briefly review the black-D$3$ brane solution. It starts with solving the ten-dimensional supergravity action in the low energy limit \cite{Horowitz:1991cd}. The bulk action is given as,
\begin{equation}
    \mathcal{S}_{10}=\frac{1}{2\kappa^2}\int d^{10}x\sqrt{-g}\left(\mathcal{R}-\frac{1}{2}\mathbf{\partial}\Phi\cdot\mathbf{\partial}\Phi-\frac{2}{5!}F^{2}\right).
\end{equation}
Here,  $2\kappa^2\sim g_s^2\ell_s^8$; $g_s$ is string theory coupling constant which is related to the $3+1$ dimensional YM theory coupling as $4\pi g_s=g_{YM}^2$ and $\ell_s$ is fundamental string length. $g$ and $R$ are the determinants of the metric and Ricci scalar defined on the $9+1$ dimensional background spacetime, $\Phi$ is the dilaton field and $F$ is the RR 5-form field. The black D$3$ brane solution \cite{Horowitz:1991cd} has been found as,
\begin{align}\label{extendedbh}
     dS^{2}=&-\left(1-\frac{r_{+}^4}{r^4}\right)\left(1-\frac{r_{-}^4}{r^4}\right)^{-\frac{1}{2}}dt^{2}+\left(1-\frac{r_{-}^4}{r^4}\right)^\frac{1}{2}dx_{i}dx^{i}\nonumber\\\quad&+\frac{dr^2}{\left(1-\frac{r_{+}^4}{r^4}\right)\left(1-\frac{r_{-}^4}{r^4}\right)}+r^{2}d\Omega_{5}^2,
\end{align}
where $(t,\,x_i)$ coordinates define the four dimensional worldvolume of the D$3$ brane and $\Omega_5$ is transverse five-sphere. $r$ defines the radial coordinate. $r_+$ and $r_-$ are respectively the positions of horizon and singularity. Now, making a coordinate transformation $ r =\rho\left(1+\frac{r_{-}^4}{\rho^4}\right)^{1/4} $, in \eqref{extendedbh}, the black D3 brane solution can be written as 
\begin{align}
     ds_{10}^2 &= \left(1+\frac{r_-^4}{\rho^4}\right)^{-\frac{1}{2}}\left(-f(\rho)\text{d}t^2+\text{d}\vec{x}^2\right)+\nonumber\\&\quad\quad\quad\quad\left(1+\frac{r_-^4}{\rho^4}\right)^{\frac{1}{2}}\left(\frac{\text{d}\rho^2}{f(\rho)}+\rho^2\text{d}\Omega_5^{2}\right),  \\&
    \text{where,}\,f(\rho)=1-\frac{\rho_h^4}{\rho^4}\hspace{1cm}\rho_h^4= r_+^4-r_-^4 ,\\
    & \text{and,}\quad \Phi=\Phi_0 \,(\text{constant}), \quad F_{[5]}=Q\left(1+\,^*\right)\text{Vol}\left(\Omega_5\right) .&
\end{align}
In this new coordinate, the metric horizon is located at $\rho=\rho_h$. The ADM mass and charge of the brane are,
\begin{equation}
    M\sim (5r_+^4-r_-^4) ,\quad\quad\quad Q^2= 4r_+^2{r_-}^2 .
\end{equation}
In the extremal or BPS limit, the solution reduces to BPS D$3$ brane. At this limit, $r_-^4=r_+^4=R_4^4\ell_s^4$, where $R_4^4=4\pi g_s N_c$. For non-extremal cases, we have $\rho_h^4=r_+^4-r_-^4$. We can see at the BPS limit, $\rho_h\to 0$ gives an extremal solution.
We will be working in non-extremal limits.

Now we take following scaling, $r_-^4\equiv R_4^4\ell_s^4$, $\rho_h\equiv r_h\ell_s^2$ and $\rho\equiv \ell_s^2 r$, where $r_h$ is a fixed point on $r$. Then we take the decoupling limit ($\ell_s\to 0$) in the metric, which gives decoupled geometry near the brane
\begin{equation}\label{dc_bg}
    \frac{ds_{10}^2}{\ell_s^2}= \frac{r^2}{R_4^2}\left(-f(r)\text{d}t^2+\text{d}\vec{x}^2\right)+\frac{R_4^2}{r^2}\left(\frac{\text{d}r^2}{f(r)}+r^2\text{d}\Omega_5^2\right),
\end{equation}
where, \begin{equation}
    f(r)=1-\frac{r_h^4}{r^4} .
\end{equation}
 This near-horizon geometry is AdS$_5$-Schwarzschild $\times$ S$_5$ which provides a finite temperature to the boundary gauge theory. The temperature is 
\begin{equation}
    T=\frac{r_h}{\pi R_4^2}=\frac{r_h}{\pi\sqrt{g_{YM}^2N_c}} .
\end{equation}
Here the bulk theory has two independent parameters -- temperature $T$ and string constant $g_s$. The holographic boundary theory is pure YM theory which also has two free parameters -- temperature $T$ and YM coupling $g_{YM}^2=4\pi g_s$. Since the transverse five-sphere has a constant radius, the background action is reducible to the five-dimensional integration as given below.
\begin{equation}\label{5d-action}
    \mathcal{S}_{bulk}=\frac{1}{g_s^2}\int R_4^5\text{d}\Omega_5\int\text{d}^5x\sqrt{-g_{(5)}}\left(\mathcal{R}_{(5)}+\frac{12}{R_4^2}\right),
\end{equation}
and further, the Einstein equation can be written for the five-dimensional AdS background as,
\begin{equation}
    \mathcal{R}_{(5)\mu\nu}-\frac{1}{2}\left(\mathcal{R}_{(5)}+\frac{12}{R_4^2}\right)g_{(5)\mu\nu}=0 ,
\end{equation}
where $\mathcal{R}_{(5)\mu\nu}$ and $\mathcal{R}_{(5)}$ are the Ricci tensor and Ricci scalar respectively for the five-dimensional AdS background metric $g_{(5)\mu\nu}$ with the AdS radius $R_4$. 

For the upcoming pole-skipping calculations, we need to frame the background solutions in the Eddington-Finkelstein (EF) coordinate. Our present background \eqref{dc_bg} can be transformed into ingoing EF coordinate with the transformation $v=t+\int\frac{R_4^2\text{d}r}{r^2f(r)}$. In EF coordinate, the metric takes the form,
\begin{align}
    \frac{ds_{10}^2}{\ell_s^2}&= \frac{r^2}{R_4^2}\left(-f(r)\text{d}v^2+\text{d}\vec{x}^2\right)+2\text{d}v\text{d}r+R_4^2\text{d}\Omega_5^2\nonumber\\&=g_{(5)\mu\nu}\text{d}x^\mu\text{d}x^\nu+R_4^2\text{d}\Omega_5^2 .
\end{align}
Here, $v$ is the null coordinate, $r$ is the radial coordinate, and $\vec{x}$ corresponds to the spatial coordinate $x,y,z$. Now, using the standard polar coordinates on the $S_5$, we can be written as,
 \begin{equation}
     \text{d}\Omega_5^2 =\text{d}\theta^2+\sin^2\theta\text{d}\Omega_3^2+\cos^2\theta\text{d}\varphi^2.
 \end{equation}

\subsection{Embedding D7 brane }
Without disturbing the background geometry, the transverse hypersphere can be unfolded as
\begin{align}
  \frac{ds_{10}^2}{\ell_s^2}=&\frac{r^2}{R_4^2}\left(-f(r)\text{d}v^2+\text{d}\vec{x}^2\right)+2\text{d}v\text{d}r+R_4^2\left[\text{d}\theta^2\nonumber\right.\\&\left.+\sin^2\theta\text{d}\Omega_3^2+\cos^2\theta\text{d}\varphi^2\right] ,   
\end{align}
where the three sphere $\Omega_3$ is spanned by the angular coordinates $\{\vartheta^1,\,\vartheta^2,\,\vartheta^3\}$. The D$3$ brane is delocalised along $\{v,\,x,\,y,\,z\}$ while localised in $r$ and $\Omega_5$. We consider D$7$ brane as a probe brane in this background. This D$7$ brane is placed in such a way that it fills the D$3$ brane's world volume and further extends along $\{r,\,\Omega_3\}$. Again, this D$7$ brane is wrapped on $\Omega_3$, hence it is localised in $\{\theta,\,\varphi\}$. 

In Poincare coordinate it can be easily shown that the six-dimensional transverse space $\mathbb{R}^1\times\mathbb{S}^5$ (given by $\{r,\,\Omega_5\}$) of D$3$ brane has been separated here into two part as $\mathbb{R}^1\times\mathbb{S}^3$ (given by $\{r\sin\theta,\,\Omega_3\}$) and $\mathbb{R}^1\times\mathbb{S}^1$ (given by $\{r\cos\theta,\,\varphi\}$). So the angle between $\mathbb{S}^3$ and $\mathbb{S}^1$ can be found to be $\theta$. We parameterize the localization of D$7$ in terms of embedding function $\theta\equiv\theta(r)$ and $\varphi=\text{constant}$ on $\mathbb{R}^1\times\mathbb{S}^1$ surface. Due to the rotational symmetry on $\mathbb{S}^1$ of $\mathbb{R}^1\times\mathbb{S}^1$, the distance of the D$7$ brane from the D$3$ brane is the radius of $\mathbb{S}^1$, $r\cos\theta(r)$.  

The parameterization of the eight-dimensional world volume of D$7$ brane in EF coordinates is as follows.
\begin{eqnarray}
    & \xi^0\equiv v;\quad \xi^i\equiv x^i;\,\,\text{for }i=1,\,2\,\&\,3; \quad \xi^4\equiv r; & \\
    & \xi_{\Omega_3}^i\equiv \vartheta^i,\,\,\text{for }i=5,\,6\,\&\,7\,;\theta\equiv \theta(r)\,, \varphi=\text{constant}. 
\end{eqnarray}

\begin{table}[t]
\centering
\caption{AdS$_5$ is covered by co-ordinate ${t,\vec{x},r}$ and the $S^5$ is covered by ${\vartheta_1,\vartheta_2,\vartheta_3},\theta$ and $\varphi$}
\begin{tabular}{|c|c|c|c|c|c|c|c|c|c|c|}
 \hline 
& $t$ & $x$ & $y$  & $z$ & $r$ & $\vartheta_1$ & $\vartheta_2$ & $\vartheta_3$ & $\theta$ & $\varphi$ \\ 
 \hline
 D3 & $\checkmark$ & $\checkmark$ & $\checkmark$ & $\checkmark$ & $\times$ & $\times$ & $\times$ & $\times$ & $\times$ & $\times$\\
 \hline
D7 & $\checkmark$ & $\checkmark$ & $\checkmark$ & $\checkmark$ & $\checkmark$ & $\checkmark$ & $\checkmark$ & $\checkmark$ & $\times$ & $\times$\\
 \hline
\end{tabular}
\end{table}
The induced metric on the D$7$ brane is expressed as, 
\begin{align}
    \frac{ds_{D7}^2}{\ell_s^2}  =& g_{(5)vv}\text{d}v^2+2g_{(5)vr}\text{d}v\text{d}r +R_4^2\theta'(r)^2\text{d}r^2 \nonumber\\&+ g_{(5)x^ix^i}\text{d}x^i\text{d}x^i+R_4^2\sin^2\theta(r)\text{d}\Omega_3^2, \nonumber\\
     =& \gamma_{(5)\alpha\beta}\text{d}\xi^\alpha\text{d}\xi^\beta +R_4^2\sin^2\theta(r)\text{d}\Omega_3^2, \nonumber\\
     = & \gamma_{(8)\alpha\beta}\text{d}\xi^\alpha\text{d}\xi^\beta .\label{gamma8}
\end{align}
The induced metric components in EF coordinates are 
\begin{align}\label{gamma5}
     \gamma_{(5)\a\b} =&  g_{(5)\mu\nu}\frac{\text{d}x^{\mu}}{\text{d}\xi^{\a}}\frac{\text{d}x^{\nu}}{\text{d}\xi^{\b}}+R_4^2\theta'(r)^2\delta_{\a}^r\delta_{\b}^r \nonumber\\=& g_{(5)\mu\nu}\delta_{\a}^{\mu}\delta_{\b}^{\nu}+R_4^2\theta'(r)^2\delta_{\a}^r\delta_{\b}^r .
\end{align}
The DBI action in the D$7$ brane worldvolume is,
\begin{equation}\label{8dbi_action}
    \mathcal{S}_{D7} = -N_fT_7\int \text{d}v\text{d}r\text{d}^3x\text{d}^3\vartheta\times\ell_s^8\sqrt{-\gamma_{(8)}},
\end{equation}
where, $\gamma_{(8)}=\text{Det}\left[\gamma_{(8)\alpha\beta}\right]$ is the determinant of the eight dimensional worldvolume metric \eqref{gamma8} of the D$7$ brane. $T_7=1/(g_s\ell_s^8)$ is the tension of the D$7$ brane. As $\Omega_3$ has the same symmetries as $\Omega_5$ of the D$3$ brane background, these three spheres can be integrated out easily and the DBI action can be reduced into the following form,
\begin{equation}
    \mathcal{S}_{D7} = -\frac{N_fR_4^3V_{S^3}}{g_s}\int \text{d}^5x\sin^3\theta(r)\sqrt{-\gamma_{(5)}} ,
\end{equation}
where, $\gamma_{(5)}=\text{Det}\left[\gamma_{(5)\alpha\beta}\right]$ is the determinant of the five dimensional metric given in \eqref{gamma5}. The total bulk action in presence of $N_f$ D$7$ flavor brane ($N_f\ll N_c$), 
\begin{align}
    \mathcal{S}_{total} = & \mathcal{S}_{bulk}+\mathcal{S}_{D7}, \nonumber\\
     = & \frac{1}{g_s^2}\int \text{d}^5x\sqrt{-g_{(5)}}\left[V_{S^5}R_4^5\left(\mathcal{R}_{(5)}+\frac{12}{R_4^2}\right)\right.\nonumber\\&\left.\quad\quad\quad-g_sN_fV_{S^3}R_4^3\sin^3\theta(r)\frac{\sqrt{-\gamma_{(5)}}}{\sqrt{-g_{(5)}}}\right],\nonumber\\
     = & \frac{V_{S^5}R_4^5}{g_s^2}\int \text{d}^5x\sqrt{-g_{(5)}}\left[\left(\mathcal{R}_{(5)}+\frac{12}{R_4^2}\right)\right.\nonumber\\&\left.\quad\quad-\frac{n_f}{R_4^2}\sin^3\theta(r)\frac{\sqrt{-\gamma_{(5)}}}{\sqrt{-g_{(5)}}}\right],
\end{align}
where, $V_{S^3}=2\pi^2$ and $V_{S^5}=\pi^3$ and, 
\begin{equation}
n_f=\frac{V_{S^3}}{V_{S^5}}g_sN_f =\frac{V_{S^3}}{4\pi V_{S^5}}R_4^4\frac{N_f}{N_c}\ll 1. 
\end{equation}
We call $n_f$ the probe parameter or the flavor parameter as it is only proportional to the number of flavor $N_f$ in YM theory. As we are in the probe limit, i.e., $n_f\ll 1$, this parameter is considered as the perturbation parameter. Since this total action is invariant under the variation of the five-dimensional background metric $g_{(5)\mu\nu}$, we find Einstein's equation
\begin{equation}\label{ein_eq}
    \mathcal{R}_{(5)\mu\nu}-\frac{1}{2}\left(\mathcal{R}_{(5)}+\frac{12}{R_4^2}\right)g_{(5)\mu\nu}+\mathcal{T}_{\mu\nu}^{flavor}=0.
\end{equation}
The energy-momentum tensor due to the D$7$ brane is given as follows.
\begin{align}\label{tmunu}
    \mathcal{T}_{\mu\nu}^{flavor} =& -\frac{n_f}{R_4^2}\frac{\sin^3\theta(r)}{\sqrt{-g_{(5)}}} \frac{\delta}{\delta g_{(5)}^{\mu\nu}}\sqrt{-\gamma_{(5)}} \nonumber\\=& -\frac{n_f}{2R_4^2}\sin^3\theta(r)g_{(5)\mu\rho}g_{(5)\nu\sigma}\delta_\alpha^\rho\delta_\beta^\sigma \gamma_{(5)}^{\alpha\beta}\frac{\sqrt{-\gamma_{(5)}}}{\sqrt{-g_{(5)}}} .
\end{align}
Here $\delta_{\mu}^{\alpha}$ indicates the Dirac delta function. If we take perturbation in the background metric $g_{(5)\mu\nu}$ of \eqref{dc_bg}, the induced metric $\gamma_{(5)\alpha\beta}$ in \eqref{gamma5} is also varied accordingly.

\subsection{In the presence of Maxwell's Field}
In the presence of Maxwell field $F_{\alpha\beta}$ along $\{v,r,\vec{x}\}$ with the flavors, the DBI action \eqref{8dbi_action} in the D$7$ brane worldvolume is modified into the following form,

\begin{eqnarray}
    \mathcal{S}_{D7} & = & -N_fT_7\int \text{d}^8x\sqrt{-\text{Det}\left[\ell_s^2\gamma_{(8)\alpha\beta}+2\pi\ell_s^2F_{\alpha\beta}\right]}, \nonumber\\
  & = & -\frac{N_f}{g_s}\int \text{d}^8x\sqrt{-\text{Det}\left[\gamma_{(8)\alpha\beta}+F_{\alpha\beta}\right]}, \nonumber\\
    & = &  -g_s^{-1}N_fR_4^3V_{S^3}\int \text{d}^5x\sqrt{-\gamma_{(5)}}\sin^3\theta(r)\mathcal{L}_8 ,\label{dbi_action}
\end{eqnarray}
where $\ell_s^2\gamma_{(8)\alpha\beta}$ is the pullback of the ten-dimensional background onto the D$7$ brane worldvolume. Note that $2\pi$ factor has been absorbed in $F_{\alpha\beta}$ and
\begin{equation*}
    \mathcal{L}_8 = \left[1-\frac{1}{2}F_{\a}^{~\b}F_{\b}^{~\a}\right]^{1/2} \approx \left[1-\frac{1}{4}F_{\a}^{~\b}F_{\b}^{~\a}\right] ,
\end{equation*}
The indices of the Maxwell field are raised/lowered with the D$7$ brane metric $\gamma_{(5)\alpha\beta}$, i.e., $F_\alpha^{~\beta}=F_{\alpha\lambda}\gamma_{(5)}^{\lambda\beta}$. Now to find the energy-momentum tensor \eqref{tmunu} with the presence of the Maxwell field, we need to vary this action \eqref{dbi_action} w.r.t $g_{(5)\mu\nu}$
\begin{align}
    \frac{\Delta\mathcal{S}_{D7}}{\Delta g_{(5)\mu\nu}} =& -g_s^{-1}N_fR_4^3V_{S^3}\int \text{d}^5x\sin^3\theta(r)\sqrt{-\gamma_{(5)}}\nonumber\\&\times\left[\frac{1}{2}\left(1-\frac{1}{4}F_{\l}^{~\g}F_{\g}^{~\l}\right)\gamma_{(5)}^{\a\b}+\frac{1}{2}F^{\a\l}F_{\l}^{~\b}\right]\delta^{\mu}_{\a}\delta^{\nu}_{\b} .
\end{align}
Therefore, the energy-momentum tensor is 
\begin{align}\label{stress_tensor}
    \mathcal{T}^{\mu\nu}_{flavor} =& -\frac{n_f}{2R_4^2}\sin^3\theta(r)\frac{\sqrt{-\gamma_{(5)}}}{\sqrt{g_{(5)}}}\left[\left(1-\frac{1}{4}F_{\l}^{~\g}F_{\g}^{~\l}\right)\gamma_{(5)}^{\a\b}\nonumber\right.\\&\quad\quad\quad\left.+F^{\a\l}F_{\l}^{~\b}\right]\delta^{\mu}_{\a}\delta^{\nu}_{\b} .
\end{align}
We have seen that the effective metric on the D$7$ brane due to the Maxwell field is $\gamma_{(8)\a\b}+F_{\a\b}$. Therefore the presence of an electric field modifies the location of the singularity.

\subsection{Embedding equation}
In the holographic picture, considering Maxwell's field on the flavored D$7$ brane is equivalent to considering the charged flavored quarks in the boundary Yang-Mills theory. We put a constant electric field $E$ along the $x_1$ direction on the D$7$ brane. We can take one of the two configurations of the vector field $A_\mu$ given below which gives the same results in our case. (i) the Coulomb potential $A_v(x)=E\,x_1$. It gives the Maxwell field as,
$$ F_{vx_1}  =  -\partial_{x_1}A_v = -E,$$
Or, (ii) a $v$ dependent vector potential $A_{x_1}=-E\,v$ along $x_1$ direction. It gives the same Maxwell fields as above. Here we also add a current along $x^1$. So to apply the Maxwell potential, we can take either (i) $A_v=E x^1\,\&\, A_{x^1}=\tilde{A}_x(r)$ or (ii) $A_{x^1}=-E v+\tilde{A}_x(r)$. Now, the D$7$ brane action can be rewritten as,
\begin{align*}
    \mathcal{S}_{D7} \sim & \int dr\,\mathcal{L}. \\    \mathcal{L}  = & \frac{r^{2}\sin^3\theta(r)}{R_4^3}\left[r^2+\tilde{A}_{x}^{'}\left(-2ER_{4}^{2}+r^{2}f(r)\tilde{A}_{x}^{'}\right)\right.\\&\left.+\left(-E^2R_{4}^{4}+r^{4}f(r)\right)\theta'(r)^2\right]^{1/2}.
\end{align*}
Here prime denotes the derivative with respect to $r$. 
Taking $ E=E_0/R_4^2 $, 
the Lagrangian becomes
\begin{align}
    \mathcal{L} = &\frac{r^{2}\sin^3\theta(r)}{R_4^3}\left[r^2+\tilde A_{x}^{'}\left(-2E_0+r^{2}f(r)\tilde A_{x}^{'}\right)\nonumber\right.\\&\quad\quad\left.+\left(-E_0^2+r^{4}f(r)\right)\theta'(r)^2\right]^{1/2} .
\end{align}
 From the above Lagrangian, it is clear that $\tilde A_x(r)$ is a cyclic coordinate, and hence the associated conserved current source $J_0$ is expressed as,
\begin{align}
     &\frac{\partial\mathcal{L}}{\partial \tilde{A}_x^{'}(r)}=J_0 \Rightarrow r^2f(r)\tilde{A}_x^{'}(r)=E_0\nonumber\\&\hspace{1cm}-R_4^3J_0\sqrt{\frac{\left(r^4f(r)-E_0^2\right)\left(1+r^2f(r)\theta^{'}(r)^2\right)}{r^6f(r)\sin^6\theta-R_4^6J_0^2}} .
\end{align}
   
Using this constraint, we can write the Lagrangian in the Legendre transformed form as

\begin{align}
    \tilde{\mathcal{L}} = & \mathcal{L} - J_0 \tilde{A}_x^{'}(r), \nonumber\\
    = & \frac{1}{r^2 f(r) R_4^3} \left[ -R_4^3 E_0 J_0 + \left(\left\{ r^4 f(r) - E_0^2 \right\}\times \right.\right.\nonumber\\&\left.\left.\left\{ 1 + r^2 f(r) \theta^{'}(r)^2 \right\} \left\{ r^6 f(r) \sin^6 \theta - R_4^6 J_0^2 \right\}\right)^{1/2} \right], \nonumber\\
    = &\frac{1}{r^2 f(r) R_4^3}\left(\left\{ r^4 f(r) - E_0^2 \right\} \left\{ 1 + r^2 f(r) \theta^{'}(r)^2 \right\} \right.\nonumber\\&\left.\hspace{2cm}\times\left\{ r^6 f(r) \sin^6 \theta - R_4^6 J_0^2 \right\}\right)^{1/2}. \label{final_lag}
\end{align}
\normalsize

In the last equation, we have dropped the term $-R_4^3E_0J_0$ as it is independent of the generalized coordinates. In other words, this $\theta(r)$ independent term is associated with a constant energy contribution in the D$7$ brane worldvolume. Now the final Lagrangian \eqref{final_lag} is to be real. However, it has two factors -- $r^4f(r)-E_0^2$ and $r^6f(r)\sin^6\theta-R_4^6J_0^2$ -- which can make the Lagrangian imaginary. The Lagrangian is real if either both factors are positive or negative. At finite electric field and current, both the factors are positive at large $r\to\infty$ and they decrease as we go to the smaller $r$ regime. Then both factors become simultaneously zero at a particular point $r=r_0$. Here at this special cut-off point, the Lagrangian \eqref{final_lag} vanishes. This point is also called as the position of vanishing locus \cite{Albash:2007bk, Albash:2007bq}. After that as $r<r_0$, both factors become negative which makes the Lagrangian real again. The location of the vanishing locus is at $r=r_0$ if 
\begin{equation}\label{cond1}
    r_0^4-r_h^4-E_0^2=0,
\end{equation}
\begin{equation}\label{cond2}
    r_0^6f(r_0)\sin^6\theta(r_0)=R_4^6J_0^2.
\end{equation}

Again the effective open string metric on the probe brane is given by
\begin{align*}
    \tilde\gamma_{(8)\a\b} =&  \gamma_{(8)\a\b}+F_{\a\l}\gamma_{(8)}^{~\l\n}F_{\n\b}  \nonumber\\=&  \gamma_{(5)\a\b}+ F_{\a\l}\gamma_{(5)}^{~\l\n}F_{\n\b}+R_4^2\sin^2\theta(r)d\Omega_3^2\,,
\end{align*}
and the line element is 
\begin{widetext}
\begin{align}\label{openmetric}
    \tilde{ds}_{D7}^2 = & - \frac{r^4 f(r) - E_0^2}{R_4^2 r^2} \text{d}v^2 
    + \frac{r^4 \left( r^4 f(r) - E_0^2 \right) \sin^6 \theta}{R_4^2 \left( r^6 f(r) \sin^6 \theta - R_4^6 J_0^2 \right)} \text{d}x_1^2  + \frac{r^2}{R_4^2} \left( \text{d}x_2^2 + \text{d}x_3^2 \right) 
    + 2 \left( 1 - \frac{E_0^2}{r^4 f(r)}\right.\nonumber\\&\left. + \frac{R_4^3 J_0 E_0}{r^4 f(r)} \sqrt{ \frac{ \left( r^4 f(r) - E_0^2 \right) \left( 1 + r^2 f(r) \theta^{'}(r)^2 \right)}{ r^6 f(r) \sin^6 \theta - R_4^6 J_0^2 }} \right) \text{d}v \text{d}r + R_4^2 \left( \theta'(r)^2 + \left( \frac{E_0}{r^3 f(r)} \right.\right.\nonumber\\
    & \left.\left.- \frac{R_4^3 J_0}{r^3 f(r)} \sqrt{ \frac{ \left( r^4 f(r) - E_0^2 \right) \left( 1 + r^2 f(r) \theta^{'}(r)^2 \right)}{ r^6 f(r) \sin^6 \theta - R_4^6 J_0^2 }} \right)^2 \right) \text{d}r^2 + R_4^2 \sin^2 \theta \text{d}\Omega_3^2.
\end{align}
\end{widetext}
In the region $r<r_0$, according to the aforementioned discussion, the effective open string metric changes signature, as its $vv$-component becomes positive. So, it puts a further condition on the complete system and suggests to define an effective horizon at $r^4f(r)|_{r_0}=E_0^2$, i.e. $r_0=\left(r_h^4+E_0^2\right)^{1/4}$. In the following sections, this effective horizon will play a crucial role in chaos and pole-skipping. The AdS$_5$ blackhole background has the horizon at $r=r_h$. In the presence of the D$7$ brane and the Maxwell field, the background has to be back-reacted in principle. The back-reacted background is expected to have a new horizon around $r_0$. But, using the perturbative approximation ($N_f<<N_c$), we have neglected the back-reaction. So, we don't find this effective horizon in the bulk metric. The effective horizon appears in the open string effective metric on the D$7$ brane. The above two conditions \eqref{cond1} $\&$ \eqref{cond2} give
\begin{eqnarray}
    && r_0=\left(r_h^4+E_0^2\right)^{1/4},\\
    && J_0 = R_4^{-3}E_0\left(r_h^4+E_0^2\right)^{1/4}\sin^3\theta_0,\label{current}
\end{eqnarray}
and, 
\begin{equation}
    \tilde{A}_x'(r_0) = \frac{r_0^2}{E_0}\left[1-\sqrt{\frac{2r_0^4+2r_0^2E_0^2\theta'(r_0)^2}{2r_h^4+3\left(1+r_0\theta^{'}(r_0)\cot\theta(r_0)\right)E_0^2}}\right].\\\label{gauge_pot}
\end{equation}
The above equation \eqref{current} relates the current density with the applied electric field, similar to Ohm's law. For weaker field ($E_0\ll r_h^2$), $J_0\propto E_0$, which is reminiscent of linear response theory. This field-induced current is the flow of the charged quarks in the direction parallel to the electric field. Therefore, the charged quark density distribution on dual field theory is varied with the applied field $E_0$. 
\par
Looking at the Lagrangian \eqref{final_lag}, one realises that the embedding angle $\theta(r)$ of the D$7$ brane in the D$3$ brane background is the only canonical variable in the effective D7 brane Lagrangian. Here, we take $r$ as the parameter and derive the dynamical equation of $\theta(r)$ as, 
    \begin{align}
     &\frac{d}{dr}\left[\frac{\theta'(r)}{1+r^2f(r)\theta'(r)^2}\right]-\frac{3r^4\sin^5\theta\cos\theta}{r^6f(r)\sin^6\theta-R_4^6J_0^2}\nonumber\\&+\frac{1}{1+r^2f(r)\theta'(r)^2}\theta'(r)\left[\frac{2r^3}{r^4f(r)-E_0^2} +\frac{r\left(2-f(r)\right)}{1+r^2f(r)\theta'(r)^2}\nonumber\right.\\&\left.\times\left(\theta'(r)^2+r^2f(r)\theta'(r)\theta''(r)\right) +\frac{r^4\sin^5\theta}{r^6f(r)\sin^6\theta-R_4^6J_0^2}\right.\nonumber\\&\left.\times\left(r(2+f(r))\sin\theta+3r^2f(r)\theta'(r)\cos\theta\right)\right]=0 . \label{theta_eq}
\end{align}
The equation of motion of $\theta(r)$ given in \eqref{theta_eq} is a second-order non-linear differential equation. We know that $r\cos\theta(r)$($=d(r)$) represents the separation between D$3$ and D$7$ brane. $r\sin\theta(r)$($=\sqrt{r^2-d^2}$) is the radius of the $\Omega_3$ sphere on which the D$7$ is wrapped. Therefore, $d(r)$ should be regular at the horizon of the background geometry. It refers to the regularity condition of $\theta(r)$. Furthermore, the asymptotic behaviour of $\theta(r)$ is known to be holographically related to the quark mass $m_q$ and its condensate $c_q$ as follows,
\begin{equation}\label{theta_asym} 
    d(r\to\infty)\approx m_q+\frac{c_q}{r^2}\quad \text{or}\quad \theta(r\to\infty)\approx\frac{\pi}{2}-\frac{m_q}{r}-\frac{c'}{r^3} ,
\end{equation}
where, $c'=c_q+m^3/6$. Since this system preserves chiral symmetry, there is no non-zero mass with zero condensate.
\begin{figure*}[t]
\centering
\includegraphics[width=0.7\textwidth,height=5cm]{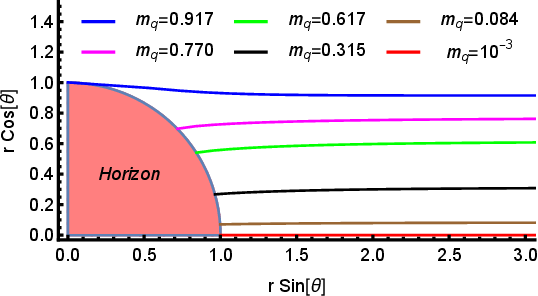}  \includegraphics[width=0.7\textwidth,height=5cm]{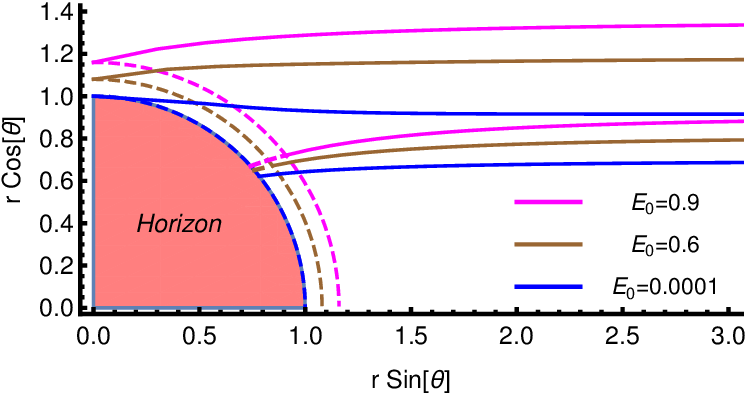}
\caption{{\it Top}: Plotted the distance between D3 brane and D7 brane - $r\cos{\theta}$ and the radius of $\Omega_3$ sphere - $r\sin{\theta}$ without electric field $E_0=0$ and $n_f=1$. {\it Below}: Plotted the distance between D3 brane and D7 brane - $r\cos{\theta}$ and the radius of $\Omega_3$ sphere - $r\sin{\theta}$ with different electric field values. The dashed lines are effective horizons at some specific values of $E_0$}
   \label{hori}
\end{figure*}

Near the point $r=r_0$ of the constant energy, the regularity condition of $\theta(r)$ helps us to expand it in the Taylor series. This immediately gives the following relation
\begin{equation}\label{theta_hor}
    \theta'(r_0) = \frac{3r_0^3\cot{\theta(r_0)}}{2\left(2r_0^4+E_0^2\right)} = \frac{3\left(r_h^4+E_0^2\right)^{3/4}}{2\left(2r_h^4+3E_0^2\right)}\cot{\theta_0},
\end{equation}
where $\theta_0$ is the value of the embedding function at $r_0$. Now, the solution of $\theta(r)$ has to be found in two steps. In the first step, putting the boundary condition \eqref{theta_hor} and the value $\theta(r_0)$, we solve the embedding function $\theta(r)$ from \eqref{theta_eq} in the range $r_0$ to $\infty$. In the next step, implementing the same boundary condition at $r_0$, we solve $\theta(r)$ from \eqref{theta_eq} in the range $r_0$ to $r_h$. In this way, we solve the embedding function for the whole range of the radial coordinate. From these solutions, we find the horizon value of the embedding function $\theta(r_h)$ depending on the boundary condition $\theta(r_0)$. We now solve numerically the equation \eqref{theta_eq} with two constraints, \eqref{theta_asym} and \eqref{theta_hor}, using the above regularity condition. Once we have found the solution of $\theta(r)$, we can easily find the distance of the D$7$ brane from the D$3$ brane, $r\cos\theta(r)$ and the radius of the $\Omega_3$ sphere on which the D$7$ brane is wrapped, $r\sin\theta(r)$. The different types of embedding of the D$7$ brane can be understood clearly from the plot of these two lengths presented in Figure \ref{hori}. In the first of the two figures, we plot $r\sin\theta$ vs. $r\cos\theta$ with zero electric field. The circular line represents the location of the black hole horizon. The different curves represent the embeddings for different quark masses. Since the background has a horizon at $r=r_h$, instead of the point $r=0$, the effective distance of the flavor brane is measured from the surface of the horizon, which is a sphere of radius $r_h$. i.e., the distance of D$7$ brane from the horizon is $r\cos\theta-r_h$. We can have three situations: - (i) $r\cos\theta-r_h<0$ signifying the fact that the D$7$ brane crosses the horizon and extends inside the horizon, i.e., {\it blackhole embedding}; (ii) $r\cos\theta-r_h=0$ signifies that D$7$ brane just touches the horizon at a single point, i.e., {\it critical embedding}; and (iii) $r\cos\theta-r_h>0$ signifies the fact that the D$7$ brane does not touch the horizon and is located above the horizon, i.e., {\it Minkowski embedding}. Here at the finite temperature, we start from the blackhole embedding and end at the critical embedding in the first plot of Figure \ref{hori}. In the other plot, we turn on the electric field. We have plotted the critical embedding at three different values of the electric field: $E_0=0.0,\,0.6\,\&\,0.9$. The plot shows that with increasing electric field values, the effective horizon moves away from the original horizon and the quark mass of the corresponding embedding increases. The quark masses for the above three values of electric fields are $m_q=0.9,\,1.1\,\&\,1.3$ accordingly. Therefore, the electric field increases the quark mass for critical embedding, i.e., the critical mass. With the electric field the type of embedding is identified with respect to $r=r_0$. So in the Minkowski embedding the D7-brane shrinks above $r_0$ and in the blackhole embedding the D7-brane crosses the vanishing locus $r=r_0$. We find two kinds of solutions in the blackhole embedding. The first kind is the solutions where the D7-branes are crossing $r=r_0$ and shrinking at a point before $r_h$. These solutions contain the conical singularity between the horizon and $r_0$. The second kind of solution is smooth. They smoothly cross the original horizon $r=r_h$ without encountering any singularity. We have found that smooth embedding occurs for the electric field value $E_0<0.86 r_h^2$. So for smooth embedding, we need to maintain an electric field smaller than this critical value $E_{\text{cr}}\approx 0.86 r_h^2$.
\begin{figure*}
    \centering
\includegraphics[width=\textwidth,height=4cm]{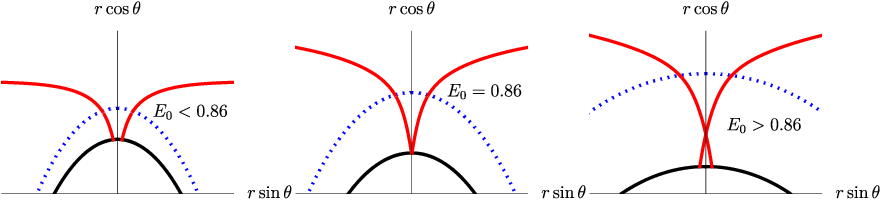}
    \caption{The plots of embedding function for different electric fields $E_0<E_{cr}$, $E_0=E_{cr}$ and $E_0>E_{cr}$ at $r_h=1$ and $\theta_0=0.1$. We find $E_{cr}=0.86$.}
    \label{fig:enter-label}
\end{figure*}

So in our study, we will consider the weak electric field to ensure the smooth embedding. For a strong electric field, the system becomes complicated due to the presence of those aforementioned singularities, and further backreaction will be important. We will work strictly in the perturbative limit.

\begin{figure*}[t]
    \centering
    \includegraphics[width=0.45\textwidth,height=5cm]{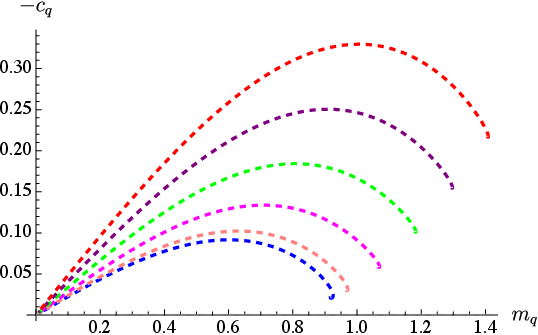}~~~~\includegraphics[width=0.45\textwidth,height=5cm]{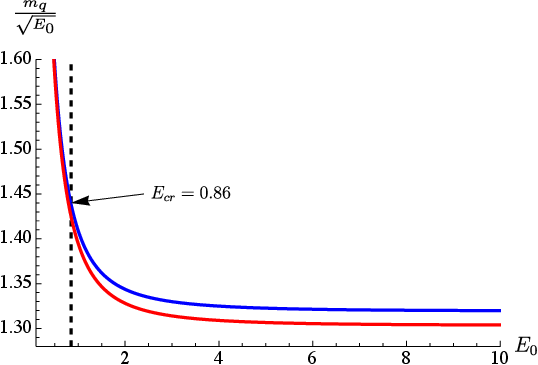}
    \caption{The condensate $c_q$ vs the quark mass $m_q$ with fixed temperature $T=1/(\pi R_4^2)$ for different electric fields. From bottom, $E_0=0.0,\,0.2,\,0.4,\,0.6,\,0.8\,\&\,1.0$. {\it Right }: Plot of $m_q/\sqrt{E_0}$ vs the electric field for two different values of embeddings.
    }
    \label{qc}
\end{figure*}

Two important meson parameters namely the quark mass $m_q$ and quark condensate $c_q$ are numerically extracted from the asymptotic expansion of $\theta$ using \eqref{theta_asym}. In Figure \ref{qc}, we have plotted those quantities for different values of the electric field $E_0$. From the plot, we find $c_q=0$ at $m_q=0$ irrespective of $E_0$. The behaviour of the quark condensate is highly non-linear with increasing mass. At low quark mass $m_q$, the magnitude of $c_q$ increases linearly. However, with increasing mass, $c_q$ reaches maximum and then starts decreasing till the point of critical mass $m_q = m^{crit}_q$, where the meson starts to melt. Therefore, beyond this mass value, the condensate does not exist. From the plot, we see that the critical mass $m^{crit}_q$ increases with the increasing value of $E_0$ \cite{Albash:2007bk, Albash:2007bq}. The binding energy of mesons is expected to be proportional to the mass of its constituent quarks. Under the influence of an external electric field, the quark and anti-quark pair of the bound state meson experience Coulomb force in opposite directions, and that helps in the dissociation of the meson by decreasing its binding energy. Hence, in a stronger electric field background, meson needs higher binding energy to survive, which leads to a higher critical mass of the quark. In previous studies \cite{Albash:2007bk}, we have also observed that this critical mass is proportional to $\sqrt{E_0}$. In Figure \ref{qc}, we have plotted $\frac{m_q}{\sqrt{E_0}}$ with $E_0$ for two different values of $\theta=0.09, 0.25$ with $r_h=1$. These embedding values are less than the critical embedding. So, with these embeddings, we will not face the conical singularity problem. At a lower value of the electric field, we can see that there is no distinction between the embeddings. The difference becomes clear at higher values of the electric field. For different $r_h$, we have calculated the critical value of the electric field numerically. For $r_{h}=0.9, E_{\text{cr}}=0.75$ and for $r_{h}=1.1, E_{\text{cr}}=0.87$. 

On the other hand, the inclusion of an electric field decreases the critical temperature of melting and the condensation parameter $c_q$ is inversely scaled with temperature. This is the reason behind the increase of $-c_q$ with $E_0$. A more detailed explanation of this model can be found in \cite{Albash:2007bk, Albash:2007bq}. 

\section{Pole-skipping and Characteristic parameters of Chaos}\label{sec_PS}
The primary goal of this work is to study the chaos in the $3+1$ dimensional flavored Yang-Mills theory. It is very common to calculate the Lyapunov exponent and butterfly velocity to measure the quantum chaos in such a model. In this context, one can get these parameters from the $vv$-component of the linearised Einstein equation. This method is well-known from various previous articles \cite{BAISHYA2024116521, Wu:2019esr}. Now, we employ this method in the above gravity background -- the embedded probe D7 branes in black D3 brane metric. We take the following perturbation in the background metric \eqref{dc_bg} 
\begin{equation}\label{pert}
    \bar{g}_{(5)\mu\nu}=g_{(5)\mu\nu}+\delta g_{\mu\nu}(r)e^{-i\omega v+i k x},
\end{equation}
 where, the perturbation is assumed to propagate along $x$ direction with frequency $\omega$ and momentum $k$. As the chaos is related to the energy density correlation function on the boundary theory, in the bulk, the corresponding perturbations need to have longitudinal polarization. These kinds of perturbation modes are also called the sound modes or scalar modes. Here, the sound modes metric perturbations are $\{\delta g_{ab},\,\delta g_{yy}+\delta g_{zz}\}$ where $a,\,b$ correspond to $v,\,r,\,x$. Without any loss of generality, we chose metric perturbations to be traceless, $\delta g_{yy}=\delta g_{zz}=-\delta g_{xx}/2$, which keeps  $\theta(r)$ invariant. Further, we consider the gauge $\delta g_{r\mu}=0$ for all $\mu$. With this, the sound channel perturbation consists of only three independent components, $\{\delta g_{vv},\,\delta g_{vx},\,\delta g_{xx}\}$.


Our main target is to look for the pole-skipping point associated with the metric perturbation variables. Therefore, we expand the metric perturbation near the effective horizon as follows,
\begin{equation*}
    \delta g_{\mu\nu}=\delta g_{\mu\nu}^{(0)}+\delta g_{\mu\nu}^{(1)} (r-r_0)+\cdots . 
\end{equation*}
Upon using these near effective-horizon expansions into the $vv$-component of the linearised Einstein equation \eqref{ein_eq}, we examine the coefficients of each order in $(r-r_0)$ expansion. The coefficients of zeroth order sound mode perturbations $\{\delta g_{vv}^{(0)},\,\delta g_{vx}^{(0)},\,\delta g_{xx}^{(0)}\}$ consists only two unknowns -- $\omega$ and $k$ resulting into following equation,

\begin{align}
     & 8 k^2 R_4^8+\left(\sqrt{2E_0^2+4 r_0^4}+2 r_0^2\right)\left(\frac{12 R_4^4 E_0^2}{r_0^4}-\frac{6 i R_4^6\omega }{r_0}\right)  \nonumber\\
        & +n_f\left[\left(\sqrt{2E_0^2+4 r_0^4}+2 r_0^2\right)\left(\frac{2 R_4^4r_0^4}{E_0^2+2 r_0^4}-2r_0^4+R_4^4\right)\right.\nonumber\\&\left.\hspace{3cm}+8 r_0^6\right]\sin^3\theta _0=0\,\\\nonumber\\&
         \text{and,}\quad\quad\quad 8 k R_4^6 \left(-3 i r_h^4+i r_0^4+R_4^2 \omega r_0^3\right) =0. \label{vv-cof}
\end{align}
      
From these equations \eqref{vv-cof}, we get, 
\begin{equation}\label{omega}
    \omega = i\frac{3r_h^4-r_0^4}{R_4^2r_0^3} = \frac{i}{R_4^2}\frac{2(\pi T R_{4}^2)^4-E_{0}^2}{\left((\pi R_{4}^2T)^4+E_{0}^2\right)^{3/4}},
\end{equation}
where, we have used the relations $r_0^4=r_h^4+E_0^2$ and $r_h=\pi R_4^2 T$.
Note that, from the right plot of Figure {\ref{qc}}, we have already observed that $E_{cr}<r_h^2$. So, the allowed value of the electric field is always $E_0<r_h^2$ which implies $E_0^2\ll T^4$. Thus the shift due to the electric field term in the effective horizon is always very small compared to the blackhole horizon.
Therefore, utilizing the relation given in \eqref{chaos}, the Lyapunov exponent is calculated to be, 
\begin{align}\label{lypunov}
    \lambda_{L} =& \frac{2(\pi T R_{4}^2)^4-E_{0}^2}{R_{4}^2\left((\pi R_{4}^2T)^4+E_{0}^2\right)^{3/4}}\,,\nonumber\\
    = & 2\pi T\left[1-\frac{5 E_0^2}{4\pi^4 R_4^8 T^4}+\frac{33 E_0^4}{32\pi^8 R_4^{16}T^8}+\mathcal{O}(E_0^5)\right]\,.
\end{align}

The Lyapunov exponent in \eqref{lypunov} is found to be affected by the electric field. The parameter $R_{4}$ is connected to the YM theory as $R_{4}^2=\sqrt{g_{\text{YM}}^{2}N_c}$. So, from this Lyapunov exponent, we can see the dependence on $N_c$ and $g_{\text{YM}}$. In the absence of an electric field, we obtain the standard expression $\lambda_{L} = 2\pi T$, which saturates the MSS bound. Therefore, without the current source, the system is maximally chaotic. From \eqref{omega}, the Lyapunov exponent is found to be zero if $r_0^4=3r_h^4$ or $E_0^2=2\pi^4R_4^8T^4$, which sets the cut-off value of field $E_0=\sqrt{2}r_h^2$. But this value is beyond the critical value $E_0^{cr}$. Further in our perturbative frame, the flavour density is too small as it is proportional to $n_f$. To keep the model consistent, the applied electric field $E_0$ should be small enough so that the induced current does not make much change in the charge distribution. So here the Lyapunov exponent can not be zero by tuning the electric field.
\begin{figure}[b]
    \centering  
    \includegraphics[width=0.4\textwidth,height=5cm]{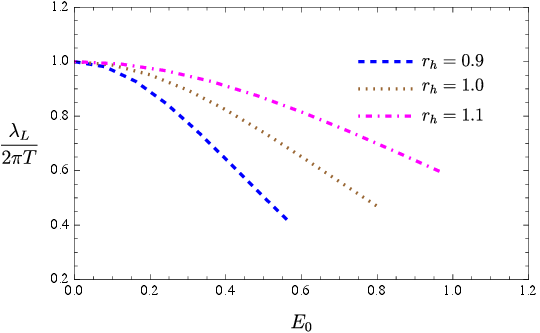}
    \caption{Plot of Lyapunov exponent $\lambda_{L}$ vs electric field $E_0$ at three different temperature $\pi R_4^2T=0.9,\,1.0\,\&\,1.1$ and $n_f=1$.}
    \label{lam}
\end{figure}

In the limit $E_{0}^{2}<<(\pi T  R_{4}^2)^4$, we can write the above Lyapunov exponent \eqref{lypunov} as,
\begin{equation}
    \lambda_{L}\sim 2\pi T\left(1-\frac{5 E_0^2}{4\pi^4 R_4^8 T^4}\right)= 2\pi T\left(1-\frac{5 E_0^2}{4\pi^4 g_{\text{YM}}^{4} T^4 N_c^2}\right).
\end{equation}

In Figure \ref{lam}, we have plotted the Lyapunov exponent $\lambda_{L}$ with the variation of the electric field. As the electric field $E_0$ increases, $\lambda_{L}$ decreases monotonically. We have plotted the $\lambda_L$ up to the critical value of the electric field only. As we have already discussed, beyond that critical value, we have conical solutions. So, we must have to avoid those solutions by restricting the electric field. Another point is, that turning on the electric field, pulls the quark/anti-quark pair apart, and lowers their binding energy. The plot, therefore, indicates an interesting fact that as meson's binding energy decreases, the system becomes less chaotic with decreasing Lyapunov exponent. 

The solution of $k$ can be found from \eqref{vv-cof} which assumes the following form, 
\begin{align}
   &  8R_4^8 k^2 = \left(\sqrt{2E_0^2+4 r_0^4}+2 r_0^2\right)\left(-\frac{12 R_4^4 E_0^2}{r_0^4}+\frac{6 i R_4^6\omega }{r_0}\right)  \nonumber\\
        & -n_f\left[\left(\sqrt{2E_0^2+4 r_0^4}+2 r_0^2\right)\left(\frac{2 R_4^4r_0^4}{E_0^2+2 r_0^4}-2r_0^4+R_4^4\right)\nonumber\right.\\&\left.\hspace{3cm}+8 r_0^6\right]\sin^3\theta _0
\end{align}
    
Inserting the solution of $\omega$ into the above equation and expanding for small $E_0$, the solution of the momentum $k$ can be written as follows,
\begin{align}
    k^2 = & -\frac{6 r_h^2}{R_4^4}-\frac{3 E_0^2}{4 R_4^4 r_h^2}\nonumber\\&+n_f \left(-\frac{r_h^2}{R_4^4}+\frac{E_0^2\left(r_h^4-3R_4^4\right)}{8r_h^2 R_4^8}\right)\sin^3\theta_0+O\left(E_0^3/R_4^8\right)
    \label{momentum}
\end{align}
where, $\theta_0\equiv\theta(r_0)$. The magnitude of momentum depends on the flavor parameter $n_f$ and the electric field $E_0$. Without the probe D$7$ brane (i.e. $n_f=0$ and $E_0=0$), we recover the standard result for pure AdS$_5$, $k^2=-6\pi^2T^2$. Unlike $\omega$, the momentum receives a contribution from $\theta_0$ which in turn leads to non-trivial quark mass $m_q$ dependence, which is one of the important results of our present study. Given the above solutions for the pole-skipping momentum, the butterfly velocity can be calculated from the relation $v_b=\lambda_{L}/|k|$ as,
     \begin{align}\label{butterfly_v}
        v_b = & \frac{r_0^2 \left(2r_h^4-E_0^2\right) \sqrt{3 E_0^2+2 r_h^4} \sqrt{\sqrt{6 E_0^2+4 r_h^4}-2 r_0^2}}{6 \sqrt{6} E_0^3 R_4^4 r_0^3 \left(3 E_0^4+8 E_0^2 r_h^4+4 r_h^8\right)^{3/2}} \nonumber\\&\times\left[-12 E_0^2 R_4^4 \left(3 E_0^4+8 E_0^2 r_h^4+4 r_h^8\right)\right.\nonumber\\&\left.+ n_f \sin^3\theta_0 \left\{4 r_0^{10}\left(2r_0^4+E_0^2\right) \sqrt{6 E_0^2+4 r_h^4} \right.\right. \nonumber\\
        & \left.\left.-12 E_0^2 r_0^{12} + E_0^4 R_4^4 r_0^4 - 2 r_0^8 \left(E_0^4-2 E_0^2 R_4^4\right) - 16 r_0^{16}\right\}\right]
    \end{align}
   
At the small electric field expansion
 \begin{align}
     v_b =& \sqrt{\frac{2}{3}}\left[1-\frac{21 E_0^2}{16\pi^4 R_4^8 T^4}+\frac{n_f\sin^3\theta_0}{12}\left(-1\nonumber\right.\right.\\&\left.\left.+\frac{E_0^2}{8R_4^4}\left(1+\frac{17}{2\pi^4R_4^4T^4}\right)\right)\right]+O\left(E_0^3/R_4^6\right) .
 \end{align}
The butterfly velocity characterises how fast information scrambles in a system in the chaotic regime, and that turns out to be a function of the electric field, mass and current density of flavors in our study. We can see the effect in terms of YM theory by replacing $R_4$ in terms of $g_{\text{YM}}$ and $N_c$. For pure AdS$_5$, the butterfly velocity is a numerical constant $\sqrt{\frac{2}{3}}$, which can be recovered from \eqref{butterfly_v}. 
\begin{figure*}
    \centering
    \includegraphics[width=0.45\textwidth,height=5cm]{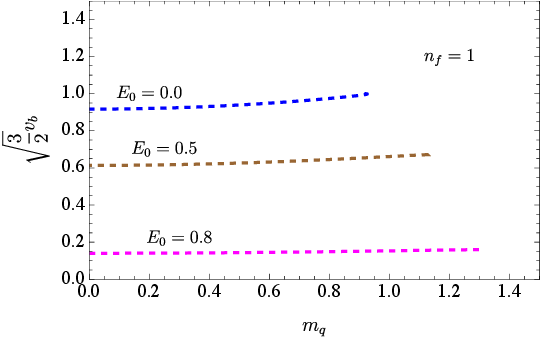}~~
~~\includegraphics[width=0.45\textwidth,height=5cm]{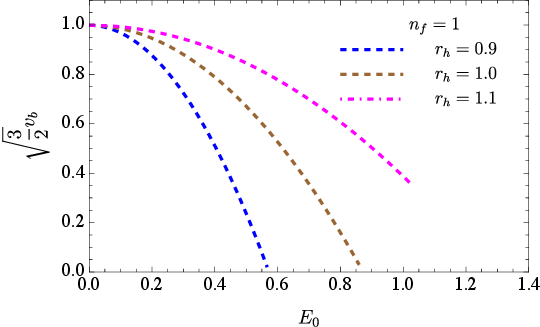}
    \caption{{\it Left}: The plot of butterfly velocity $v_b$ vs quark mass $m$ at three different fields $E_0=0.0,\,0.5\,\&\,0.8$ with fixed $T=1/(\pi R_4^2)$ and $n_f=1$. {\it Right}: The plot of butterfly velocity $v_b$ vs electric field $E_0$ at $\theta_0=10^{-2}$ at three different temperature $\pi R_4^2T=0.9,\,1.0\,\&\,1.1$ and $n_f=1$.}
    \label{vb}
\end{figure*}
In Figure \ref{vb}, we represent the effect of flavor quark on the characteristic parameters of chaos. We plot the butterfly velocity $v_b$ with the quark mass $m_q$ for three different values of electric field, $E_0=0,\,0.5\,\&\,0.8$. For a fixed value of the electric field, we find the butterfly velocity remains almost constant in the full range $0\leq m_q\leq m_q^{crit}$. However, with increasing electric field, $v_b$ decreases. The variation is almost the same as the Lyapunov exponent.

\section{Master Equation}\label{sec_master}

In the previous section, we have calculated the chaotic parameters from the $vv$ component of Einstein's equation by perturbing the sound mode. In this section, we will analyze the perturbed modes by constructing a unique master equation from all the perturbed components of Einstein's equation as shown in \cite{Kodama:2003jz}. In this D3-D7 background set-up, it would be noteworthy to study the dispersion relations at small $\omega, k$. We can study these dispersion relations from the pole-skipping of retarded two-point function of the energy-momentum tensor \cite{Blake:2018leo, Blake:2019otz, BAISHYA2024116521}. Overall, these pole-skipping points unfold the relation between the boundary Green's function in the hydrodynamic limit and low energy perturbation with its near-horizon behaviour in the bulk. We perturb the background metric as $g_{\mu\nu}+\delta g_{\mu\nu}$, from which we compute the linearised coupled Einstein's equations. However, there exists a unique choice of gauge-invariant perturbation variable, which will be shown to reduce the perturbation equations into the unique form of the second-order master equation. The near-horizon behaviour of this master equation is enough to study the hydrodynamic dispersion relations. The complete solution of the perturbations comes from the master equation of the corresponding perturbation channel. The metric perturbation variables in a generic five-dimensional black hole can be classified into three classes: scalar or sound channel, vector or shear channel and tensor channel. We have chosen the wavenumber $k$ to point along the $x$-direction. Then, we can write these perturbed channels as:
\begin{itemize}
    \item Scalar or sound channel : $\delta g_{vv},\,\delta g_{vx},\,\delta g_{xx}$ and $\delta g_{yy}+\delta g_{zz}$.
    \item Vector or shear channel : $\delta g_{vy},\,\delta g_{xy},\,\delta g_{vz}$ and $\delta g_{xz}$
    \item Tensor channel : $\delta g_{yz}$ .
\end{itemize}
We have imposed the radial gauge condition and trace-less condition, which we have thoroughly discussed in the previous section.

\subsection{Sound Channel}\label{sec_sound}
The sound channel has three independent variables $\{\delta g_{vv},\,\delta g_{vx},\,\delta g_{xx}\}$. We can construct a gauge invariant master variable from these three variables,
\begin{align*}
    \mathcal{Z}_{so}(r)=k^2\delta g_{vv}+&2\omega k\delta g_{vx}+\omega^2\delta g_{xx}\\&+\frac{k^2}{2}\left[2-f(r)-\frac{\omega^2}{k^2}\right]\left(-\delta g_{xx}\right) ,
\end{align*}
which satisfies the unique master equation as follows,
\begin{equation}\label{sound_eq}
    \mathcal{Z}_{so}''(r)+\mathcal{P}_{so}(r,\omega,k)\times\mathcal{Z}_{so}'(r)+\mathcal{Q}_{so}(r,\omega,k)\times\mathcal{Z}_{so}(r)=0 .
\end{equation}
The coefficients $\mathcal{P}_{so},\,\mathcal{Q}_{so}$ are functions of $r,\,\omega$ and $k$. The detailed expressions of these coefficients are given in appendix \ref{app_sound}. To handle these large expressions, we have expanded them around $E_0=0$ and taken up to order of $E_0^2$. These coefficients consist of the flavor correction at the order of $n_f$. As $n_f\to 0$ and $E_0\to 0$, the equation \eqref{sound_eq} reduces to the sound mode master equation of standard AdS$_5$. To obtain the pole-skipping points, we expand the master variable near the horizon as
\begin{equation*}
    \mathcal{Z}_{so}(r)= \sum_{n=0}^\infty Z_n\times (r-r_0)^n =Z_0+Z_1\times (r-r_0)+\cdots ,
\end{equation*}
where $Z_n$ are the non-zero finite constant coefficients. Consequently, the master equation around the horizon $r=r_0$ takes the following form, 
\begin{equation}\label{sound_srs}
\sum_{q=0}^\infty\left[\sum_{p=0}^{q+1}C^q_p\left(\omega,\,k\right)\times Z_p\right]\times (r-r_0)^q=0 ,
\end{equation}
where $C^q_p$ depend on $\omega,\,k$ and other parameters of the system. At each order of $(r-r_0)$, the expression in the square bracket of \eqref{sound_srs} should vanish. It gives the relation between various $Z_p$. We can write it in the well-known matrix equation as discussed in \cite{Ceplak:2019ymw, Natsuume:2019xcy}.

In this paper we discuss only the first-order pole-skipping point $(\omega_1,k_1)$, which comes from the equation \eqref{sound_srs} at the zeroth order term, i.e., $q=0$, which gives
\begin{equation*}
    C^0_0(\omega,\,k)\times Z_0+C^0_1(\omega,\,k)\times Z_1=0 .
\end{equation*}
Since, $Z_0$ and $Z_1$ are non-zero and arbitrary, we must have $C^0_1(\omega_1,\,k_1)=0$ and $C^0_0(\omega_1,\,k_1)=0$. From the expansion of master equation \eqref{sound_eq} we get two equations given below,
\begin{subequations}
\label{1st_ps_eqn}
\begin{align}
    & r_h^2\left(2 r_h-i R_4^2 \omega_1\right)+\frac{E_0^2}{r_h^2}\left\{\left(r_h-i R_4^2 \omega_1\right)+\omega_1^2\frac{3\left(2r_h-iR_4^2\omega_1\right)}{2\left(2 k_1^2-3 \omega_1^2\right)}\right\}\nonumber\\&\hspace{6cm}=0 , \label{eqn_w}\\\nonumber\\
    \hspace{-3cm}& 6r_h^2-R_4^4k_1^2+r_h\left(2r_h+iR_4^2\omega_1\right)\frac{2k_1^2+9\omega_1^2}{2k_1^2-3\omega_1^2}\nonumber\\&+n_f\frac{2 R_4^4 r_h^2 \left(14 k_1^2-9 \omega_1 ^2\right)+r_h^6 \left(10 k_1^2+9 \omega_1^2\right)}{12 R_4^4 \left(2 k_1^2-3 \omega_1^2\right)}\sin^3\theta_0\nonumber\\
    & +\frac{E_0^2}{4r_h^3\left(2 k_1^2-3 \omega_1^2\right)^2}\left[\left(32 r_h \left(8 k_1^4-9 k_1^2 \omega_1^2\right)+i R_4^2 \omega_1  \left(92 k_1^4\right.\right.\right.\nonumber\\&\left.\left.\left.-108 k_1^2 \omega_1^2+27 \omega_1^4\right)\right)\right]+\mathcal{O}(n_f^2) =0\,.\label{eqn_k}
\end{align}
\end{subequations}
Solving the above two equations simultaneously, we can evaluate the value of $\omega_1$ and $k_1$. Here, we have assumed $2k_1^2-3\omega_1^2\neq 0$ in evaluating these solutions.
 
Without D$7$ probe branes, $n_f=0$ and $E_0=0$, the first-order pole-skipping point is located at $\omega_1=-2ir_h/R_4^2$ and $k_1$ is given by the equation
\begin{equation}\label{sound_unpert}
    6r_h^2-R_4^4k_1^2+4r_h^2\frac{R_4^4k_1^2-18r_h^2}{R_4^4k_1^2+6r_h^2}=0\Rightarrow k_1^2=\frac{2r_h^2}{R_4^4}\left(1\pm 2i\sqrt{2}\right) .
\end{equation}
This solution matches with the known result obtained in \cite{Blake:2019otz, Wu:2019esr}. This result \eqref{sound_unpert} also ensures the validity of our calculations.

With D$7$ brane (i.e., $n_f\neq 0$ and $E_0=0$), the first order pole-skipping occurs at $\omega_1=-2ir_h/R_4^2$ and the equation of $k_1$ is
\begin{align*}
    &-12 R_4^4 \left(-4 k_1^2 R_4^4 r_h^2+36 r_h^4+k_1^4 R_4^8\right)\nonumber\\&+n_f r_h^2 \sin^3\theta _0\left(5 k_1^2 R_4^4 r_h^4+36 R_4^4 r_h^2-18 r_h^6+14 k_1^2 R_4^8\right)=0,
\end{align*}
from which $k_1$ can be solved and we get, 

\begin{align}
k_{1}^{2} & =\frac{1}{24 R_{4}^{12}}\left(5 n_{f}R_{4}^{4}r_{h}^{6}\sin^3\theta_{0}+2R_{4}^{8}r_{h}^{2}(24+7n_{f}\sin^{3}\theta_{0})\right.\nonumber\\&\left.\pm\left(R_{4}^{8}r_{h}^{4}(-18432 R_{4}^{8}+384 n_{f}R_{4}^{4}(8R_{4}^{4}-r_{h}^{4})\sin^3\theta_0\right.\right.\nonumber\\&\left.\left.+n_{f}^{2}(14 R_{4}^{4}+5r_{h}^{4})^{2}\sin^6\theta_0)\right)^{1/2}\,\right).
\end{align}

We, therefore, find no effect of the flavor quarks on $\omega_1$ as long as the Yang-Mills system contains only neutral meson states. 
However, the momentum value $k_1$ acquires non-trivial correction due to the presence of meson states. As the near-horizon value $\theta_0$ appears in momentum $k_1$, we find that the momentum values are affected by the quark mass $m_q$.

With D$7$ brane and non-vanishing world volume electric field, we again can solve the above two equations \eqref{eqn_w} and \eqref{eqn_k} perturbatively in the small electric field limit and up to the first-order of $n_f$ to find the first-order pole-skipping point.  
\begin{subequations}\label{1st_ps_sol}
\begin{align}
    \omega_1  = & -\frac{2 i r_h}{R_4^2}\pm\frac{R_4^6E_0^2}{r_h^3}\left[\frac{1}{\sqrt{2} R_4^8}-\frac{n_f\sin^3\theta _0}{64 R_4^{12}}\right.\nonumber\\&\left.\left(\left(\sqrt{2}\mp i\right) r_h^4+2 \left(2 \sqrt{2}\pm i\right) R_4^4\right)\right]+\mathcal{O}\left(E_0^3\right), \\ \label{sound_w_sol}
     k_1^2  = & \frac{2\left(1 \pm 2i\sqrt{2}\right)r_h^2}{R_4^4}+\frac{n_fr_h^2\sin^3\theta_0}{24 R_4^8}\left(\left(5 \pm i \sqrt{2}\right) r_h^4\right.\nonumber\\&\left.+2 \left(7 \mp 4 i \sqrt{2}\right) R_4^4\right)+\frac{R_4^4E_0^2}{r_h^2}\left(\pm \frac{33+15i\sqrt{2}}{4R_4^8} \right.\nonumber\\
     &\left. +\frac{n_f \sin^3\theta _0 }{768 R_4^{12}}\left(\left(640\pm263 i \sqrt{2}\right) r_h^4+6 \left(-32\nonumber\right.\right.\right.\\&\left.\left.\left.\pm89 i \sqrt{2}\right) R_4^4\right)\right)+\mathcal{O}\left(E_0^3\right).
\end{align}
\end{subequations}

To this end let us remind the reader that, previously, we have seen that the boundary Green's function related to the sound channel perturbations has given the pole-skipping points that are located in the upper half of the complex $\omega$-plane. Therefore, those pole-skipping points can be directly related to the chaos parameters. However, from the master equation analysis, we get the pole-skipping points \eqref{1st_ps_sol}, that are located on the lower half of the complex $\omega$-plane. These P-S points, therefore, cannot be related to the chaos parameters. However, it shows the same kind of non-uniqueness in the associated boundary Green's function. 
In the absence of the probe brane, the absolute value $|k_1^2|=6r_h^2/R_4^4$ from \eqref{sound_unpert} and $|v_b|=\sqrt{2/3}$ are the same for both of the upper and lower half-planes.
 However, perturbation modes are found to be related to the propagation and decay of the energy energy fluctuation in the boundary theory. The real and imaginary part of $k_1$ respectively gives diffusion and decay of energy density fluctuation in $x$ direction. 
We can discuss the dispersion relation of hydrodynamic modes that pass through P-S points. We have noticed that the first-order P-S point is very close to the hydrodynamic dispersion curve i.e., we can expect this point to satisfy the relation $\omega = -i\mathcal{D}_{T}k^2$. We indeed can define an effective thermal diffusion constant from the real part of $k_1$, as, 
\begin{align}
 \mathcal{D}_{T}^{\text{eff}}  = & \frac{\Re[i\omega_1]}{\Re[k_1^2]}  =  \frac{1}{2 \pi  T}-n_f\frac{\pi^4 R_4^4 T^4+1}{48 \pi  T}\sin ^3\theta _0\nonumber\\&+ \left(\frac{3}{8 \pi ^5 R_4^8 T^5} -n_f\frac{5 \left(71 \pi ^4 R_4^4 T^4-6\right)}{3072 \pi ^5 R_4^8 T^5}\sin^3\theta _0\right)E_0^2\nonumber\\&\hspace{5cm}+\mathcal{O}\left(E_0^3\right).
 \end{align}
In the absence of the probe brane and the electric field, the thermal diffusion constant is found to be,
\begin{equation}
    \mathcal{D}_{T}=\frac{1}{2\pi T}=\frac{3}{4\pi}\frac{v_b^2}{T},
\end{equation}
which is observed to be connected to the butterfly velocity  \cite{Blake:2017qgd}. The presence of the probe brane modifies the diffusion constant in a nontrivial manner. 

\subsection{Shear Channel}\label{sec_shear}
The perturbations in the direction transverse to the propagation direction are analysed in this section. These perturbation modes as a group belong to the shear channel. On the gauge theory side, these perturbations are the fluctuation of the momentum density. Using shear channel perturbations $\delta g_{vy}(r),\,\delta g_{xy}(r),\,\delta g_{vz}(r)$ and $\delta g_{xz}(r)$, we can construct two gauge invariant variables as $\mathcal{Z}_y=k\delta g_{vy}+\omega\delta g_{xy}$ and $\mathcal{Z}_z=k\delta g_{vz}+\omega\delta g_{xz}$. By manipulating the perturbed Einstein's component equations, we can construct the master equation for these gauge-invariant variables. The general form of the master equations can be written as, 
\begin{equation}\label{shear_eq}
    \mathcal{Z}_{sh}''(r)+\mathcal{P}_{sh}(r,\omega,k)\times\mathcal{Z}_{sh}'(r)+\mathcal{Q}_{sh}(r,\omega,k)\times\mathcal{Z}_{sh}(r)=0.
\end{equation}
where, $\mathcal{Z}_{sh}(r)\equiv k\delta g_{vy}(r)+\omega\delta g_{xy}(r) \equiv k\delta g_{vz}(r)+\omega\delta g_{xz}(r)$. The coefficients are given in appendix \ref{app_shear}. Using the same method as the sound mode, we find the pole-skipping point from the near-horizon expansion.

In the small electric field limit, the equations for the pole-skipping points are,
\begin{subequations}
\begin{equation}
    2 r_h^2 \left(2 r_h-i R_4^2 \omega_1\right)+\frac{E_0^2 \left(4 r_h \left(k_1^2+2 \omega_1^2\right)-i R_4^2 \omega_1^3\right)}{\omega_1^2 r_h^2}=0 ,
\end{equation}
and,
\begin{align}
        & -k_1^2 R_4^4+8 r_h^2-\frac{i R_4^2 r_h \left(4 k_1^2+7 \omega_1^2\right)}{\omega_1}+\frac{1}{4} n_f r_h^2 \sin ^3\theta_0 \nonumber\\&\left(2-\frac{r_h^4}{R_4^4}\right)  +E_0^2\left[\frac{1}{4 \omega_1^3 r_h^3}\left(32 \omega_1 r_h \left(k_1^2+\omega_1^2\right)-i R_4^2 \left(16 k_1^4\right.\right.\right.\nonumber\\&\left.\left.\left.-12 k_1^2 \omega_1^2+7 \omega_1^4\right)\right)  +\frac{n_f \sin ^3\theta_0}{8 R_4^4 \omega_1^2 r_h^2 \left(r_h^4+2 R_4^4\right)} \left(4 k_1^4 \left(r_h^4\right.\right.\right.\nonumber\\&\left.\left.\left.-2 R_4^4\right)^2+\omega_1^2 \left(-4 R_4^4 r_h^4-3 r_h^8+4 R_4^8\right)\right)\right]=0.
\end{align}
We can find the first-order pole-skipping point from the above equations. The approximate solutions in terms of the small $n_f$ limit of the above equations are:
\begin{align}
    \omega_1 & \approx  -\frac{2 i r_h}{R_4^2}-in_fE_0^2 \frac{2R_4^4-r_h^4}{8 r_h^3 R_4^6}\sin^3\theta_0+\mathcal{O}\left(E_0^3\right),\label{shear_w}\\
    k_1^2 & \approx  \frac{6 r_h^2}{R_4^4}-n_fr_h^2\frac{2 R_4^4- r_h^4}{4 R_4^8}\sin^3\theta_0+\frac{E_0^2}{r_h^2}\left[\frac{69}{2 R_4^4}\right.\nonumber\\&\left.+n_f\frac{15 r_h^8-20 R_4^4 r_h^4-4 R_4^8}{8 R_4^8 \left(r_h^4+2 R_4^4\right)}\sin^3\theta_0\right]+\mathcal{O}\left(E_0^3\right).\nonumber\\ \label{shear_k}
\end{align}
In the absence of the flavor brane and the electric field, the first-order pole-skipping point matches with the previous result $\omega_1=-2i\pi T,\,k_1^2=6\pi^2T^2$ given in \cite{Blake:2019otz, Wu:2019esr}.
From the above result, we can see the presence of probe D7 brane parameters.

Like the sound channel, the first-order pole-skipping point follows the hydrodynamic dispersion relation $\omega=-i\mathcal{D}_pk^2$ associated with the momentum transportation. Therefore, the corresponding momentum diffusion constant can similarly be calculated up to order $E_0^2$ as given below, 
\begin{align}
    \mathcal{D}_p^{\text{eff}} & =  \frac{i\omega_1}{k_1^2} \nonumber\\
    & =  \frac{1}{3\pi T}+n_f\frac{2 - \pi^4R_4^4T^4}{72 \pi T}\sin^3\theta_0-\frac{E_0^2}{\pi^5R_4^8T^5}\left[\frac{23}{12}\right.\nonumber\\&\left.+n_f\frac{76-20 \pi^4R_4^4T^4 - 5 \pi^8 R_4^8 T^8}{144 \left(2+ \pi^4R_4^4T^4\right)}\sin^3\theta_0\right].\nonumber
    \label{mom_diff}
\end{align}
In the absence of the probe brane and the electric field, the momentum diffusion constant is,
\begin{equation}\label{mom_diff_0}
    D_{p}=\frac{1}{3\pi T}=\frac{1}{2\pi}\frac{v_b^2}{T}.
\end{equation}
\end{subequations}
It is very tempting to connect the momentum diffusion constant with the butterfly velocity \cite{Grozdanov:2018kkt}. However, in a physical sense, this relationship is not very well-motivated. A factor of $3/4$ is missing in the diffusion constant calculated from the P-S point, whereas in the boundary theory, it comes out to be $\frac{1}{4\pi\, T}$ \cite{Policastro:2002se}. The presence of the probe brane modifies the diffusion constant up to higher orders of the electric field.

\subsection{Tensor channel}\label{sec_tensor}
Since the perturbation modes are assumed to propagate along the x-direction, the tensor mode will have polarization in the $y-z$ plane. Hence, $\delta g_{yz}(r)$ is the only perturbed component in this channel. Assuming $\mathcal{Z}_{ten}=\delta g_{yz}$, the master equation can be written in the following form,
\begin{equation}\label{tensor_eq}
\mathcal{Z}_{ten}''(r)+\mathcal{P}_{ten}(r,\omega,k)\times\mathcal{Z}_{ten}'(r)+\mathcal{Q}_{ten}(r,\omega,k)\times\mathcal{Z}_{ten}(r)=0 .
\end{equation}
The coefficients are given in appendix \ref{app_tensor}. Using the same method as the sound mode, we find the pole-skipping point from the near-horizon expansion.

The first-order pole-skipping point $(\omega_1,\,k_1)$ comes from the equations:
\begin{subequations}
    \begin{equation}
        4 r_h \left(E_0^2+r_h^4\right)-i R_4^2 \omega_1  \left(E_0^2+2 r_h^4\right) = 0 ,
    \end{equation}
    and,
    \begin{align}
       & 8 k_1^2 R_4^8+\frac{6 i R_4^6 \omega_1  \left(E_0^2+4 r_h^4\right)}{r_h^3}-\frac{n_f \sin^3\theta_0}{r_h^2}\left(2 R_4^4 \left(E_0^2+2 r_h^4\right)\right.\nonumber\\&\left.\quad\quad\quad\hspace{2cm}-3 E_0^2 r_h^4-2 r_h^8\right)= 0 .
    \end{align}
\end{subequations}
We, therefore, find the first-order pole-skipping point at 
\begin{subequations}
    \begin{equation}\label{w_tensor}
        \omega_1 = -\frac{4ir_h}{R_4^2}\frac{r_h^4+E_0^2}{2r_h^4+E_0^2} ,
    \end{equation}
    and,
    \begin{align}\label{k_tensor}
       k_1^2=-\frac{3\left(4r_h^4+E_0^2\right)\left(r_h^4+E_0^2\right)}{r_h^2R_4^4\left(2r_h^4+E_0^2\right)}&+\frac{n_f\sin^3\theta_0}{8r_h^2R_4^8}\left(2 R_4^4 \left(E_0^2+2 r_h^4\right)\right.\nonumber\\&\left.-3 E_0^2 r_h^4-2 r_h^8\right) .
    \end{align}
In the absence of the flavor brane, we recover the result of pure AdS$_5$ \cite{Blake:2019otz, Wu:2019esr} from the above results \eqref{w_tensor} $\&$ \eqref{k_tensor} by putting $n_f=0$ and $E_0=0$. This point is given as $\omega_1=-{2ir_h}/{R_4^2}$ and $k_1^2=-{6r_h^2}/{R_4^4}$.

The results for the neutral D$7$ brane is again can be derived from the above results with $E_0=0$. It is given as
\begin{equation}
    \omega_1=-\frac{2ir_h}{R_4^2},\quad\quad\text{and,}\quad k_1^2=-\frac{6r_h^2}{R_4^4}+n_f\sin^3\theta_0\frac{r_h^2\left(2R_4^4-r_h^4\right)}{4R_4^8} .
\end{equation}
Whereas we have found the value of $(\omega_1,\,k_1)$ for charged D$7$ brane as given in \eqref{w_tensor} and \eqref{k_tensor}. These results can be written with the small $E_0$ approximation as below,
\begin{align}
      \omega_1 = &-\frac{2ir_h}{R_4^2}-\frac{iE_0^2}{r_h^3R_4^2}+\mathcal{O}\left(E_0^3\right) ,&\\\nonumber
     \text{and,}\quad k_1^2=&-\frac{6r_h^2}{R_4^4}+n_f\sin^3\theta_0\left(\frac{r_h^2\left(2R_4^4-r_h^4\right)}{4R_4^8}\right.\nonumber\\&\left.\hspace{1cm}+\frac{2R_4^4-3r_h^4}{8r_h^2R_4^8}E_0^2\right)+\mathcal{O}\left(E_0^3\right).
\end{align}
\end{subequations}
From the above result, we can conclude that, in the presence of the charged probe D7 brane, we get the effect of the effective horizon on the pole-skipping points. Momentum value is affected by the
near-horizon (effective) value of the radius of $\Omega_3$ sphere.

\section{Discussions}\label{conclusions}
In this paper, we have studied the D3-D7 brane system in the supergravity limit, where the dual theory is a strongly coupled Yang-Mills gauge theory at finite temperature with light quark flavors. In bulk, a stack of D3 branes creates the black hole gravitational background, while D7 branes are considered to be probing this background with an electric field on their world volume. This type of boundary theory can be viewed as a model for unquenched QCD. Within this framework, we have computed pole-skipping points which are claimed to be connected with quantum chaotic parameters on the boundary theory \cite{Grozdanov:2020koi}. Specifically, we have calculated two well-known characteristic parameters of chaotic systems: the Lyapunov exponent and the butterfly velocity. Additionally, we have examined how these parameters depend on external electric fields and quark masses.
Another diagnostic to study the chaotic properties in both bulk/boundary theories is OTOC. The computation of OTOC in this set-up would be very interesting. 
In \cite{Banerjee:2018kwy}, the authors have computed OTOC in the D3-D5 brane set-up and argued that the Lyapunov exponent would increase in the presence of an electric field. In a perturbative limit, we have observed that the electric field is decreasing the chaotic exponent up to some critical value of the electric field. The OTOC computation in D$3$-D$7$ brane systems in the presence of an external electric or magnetic field is beyond the scope of our present paper, and we are presently working on this important issue.

In the absence of the electric field, we have found that the Lyapunov exponent is $2\pi T$, which indicates a maximally chaotic system similar to the flavor-less theory. However, the butterfly velocity has an implicit dependence on the flavor mass. The presence of an electric field deviates the system from maximal chaos. As one increases the electric field, the Lyapunov exponent decreases, and that has been displayed in Figure \ref{lam}. We have calculated a bound of the electric field up to which the Lyapunov exponent decreases. Beyond that value, solutions are not regular. So, this study is valid up to that critical value of the electric field. We further evaluated the butterfly velocity and presented it pictorially in Figure \ref{vb}. Analytically, we have found in \eqref{butterfly_v} that $v_b$ depends on both the quark mass (through $\theta_0$) and the electric field. We have observed that the butterfly velocity is insensitive to the quark mass but changes significantly in terms of the background electric field as expected due to its effect on the binding energy of the quark/anti-quark bound state. The most interesting observation is that there exists a critical electric field up to which both $(\lambda_{L}, v_b)$ decreases indicating the fact that the system under consideration becomes less chaotic for a sufficiently strong electric field. Due to strong pull under the electric field $(E > E^*)$, the meson state dissociates into two separate charged quarks. In the presence of a strong electric field, we have a system of Yang-Mills gauge fields and dissociated quarks showing less chaotic behaviour than the system with zero electric fields. It would be interesting to study this further.

 We have further analyzed the first-order pole-skipping points for gauge invariant gravitational perturbation using the master equations for the sound, shear, and tensor channels. We have calculated the pole-skipping points in the small external electric field limit. It was argued in \cite{Wu:2019esr, Natsuume:2019vcv, BAISHYA2024116521} that in the perturbative limit, $\omega$ values generically do not receive any contribution. However, our analysis reveals that this is indeed not true. For stringy brane systems, even in the probe limit the world volume electric field can give rise to non-trivial correction to the pole-skipping frequency $\omega$, and this is precisely due to the appearance of effective horizon outside the actual horizon of the black D3 brane. Pole-skipping points associated with the gauge invariant gravitational perturbations in the bulk correspond to dual hydrodynamic dispersion relation. Using this dual interpretation, we further computed the associated modified hydrodynamic transport coefficients, namely, thermal diffusion ${\cal D}_{ T}$ and momentum diffusion ${\cal D}_p$ constant. However, our entire analysis is within the probe limit. It would be interesting to investigate pole-skipping phenomena considering the back-reaction of those probe branes \cite{Bigazzi:2013jqa}.

\acknowledgments

BB would like to acknowledge the MHRD, Govt. of India for providing the necessary funding and fellowship to pursue research work. KN would like to thank IIT Guwahati for its hospitality during the initial phase of this work. We would like to thank all the great minds working in this field.

\onecolumngrid
\appendix
\begin{appendix}
\section{Coefficients of Master Equations: sound mode}\label{app_sound}
The coefficients of master equation \eqref{sound_eq} can be expressed as,
\begin{subequations}
\begin{eqnarray}
    \mathcal{P}_{so}(r,\omega,k) & = & P_0(r,\omega,k)+n_f P_1(r,\omega,k),\\
    \mathcal{Q}_{so}(r,\omega,k) & = & Q_0(r,\omega,k)+n_f Q_1(r,\omega,k).
\end{eqnarray}

Now,

\begin{eqnarray}
    && P_0 = \frac{-f(r) \left(3 r \omega^2+2 k^2 \left(r+i R_4^2 \omega \right)\right)+9 k^2 r f(r)^2+2 \left(2 k^2-3 \omega ^2\right) \left(2 r-i R_4^2 \omega \right)}{r^2 f(r) \left(k^2 f(r)+2 k^2-3 \omega ^2\right)},\nonumber \\\\
    && Q_0 = \left[-k^2 f(r) \left(k^2 R_4^4+32 r^2+11 i r R_4^2 \omega \right)+16 k^2 r^2 f(r)^2-2 k^4 R_4^4+k^2 \left(16 r^2 \right.\right.\nonumber\\
    &&\left.\left. +2 i r R_4^2 \omega +3 R_4^4 \omega ^2\right)+9 i r R_4^2 \omega^3\right]/\left\{r^4 f(r) \left(k^2 f(r)+2 k^2-3 \omega^2\right)\right\},
\\
    P_1 & = & 2 i k^2 \sin^3\theta(r)\left[k^4 \left(18 E_0^2 r^5 f(r)^4-3 r^4 f(r)^3 \left(10 E_0^2 r+3 i E_0^2 R_4^2 \omega+24 r R_4^4\right)+2 f(r)^2 \left(6 R_4^4 \right.\right.\right.\nonumber\\
    &&\left.\left.\left. \left(3 E_0^2 r+10 r^5\right)+15 E_0^2 r^5-7 i E_0^2 r^4 R_4^2 \omega +2 i E_0^2 R_4^6 \omega\right)+f(r) \left(-12 R_4^4 \left(5 E_0^2 r+4 r^5\right) \right.\right.\right.\nonumber\\
    &&\left.\left.\left. -30 E_0^2 r^5+4 i E_0^2 r^4 R_4^2 \omega +4 i E_0^2 R_4^6 \omega \right)+4 E_0^2 \left(3 r^5-2 i r^4 R_4^2 \omega +6 r R_4^4-2 i R_4^6 \omega \right)\right) \right.\nonumber\\
    && -3 k^2 \omega ^2 \left(5 E_0^2 r^5 f(r)^4+r^4 f(r)^3 \left(-17 E_0^2 r+2 i E_0^2 R_4^2 \omega-20 r R_4^4\right)+f(r) \left(-2 R_4^4 \left(17 E_0^2 r \right.\right.\right.\nonumber\\
    &&\left.\left.\left.\left. +24 r^5\right)-17 E_0^2 r^5+10 i E_0^2 r^4 R_4^2 \omega+4 i E_0^2 R_4^6 \omega\right)+f(r)^2 \left(2 R_4^4 \left(5 E_0^2 r+34 r^5\right)+17 E_0^2 r^5 \right.\right.\right.\nonumber\\
    &&\left.\left.\left. -2 i E_0^2 r^4 R_4^2 \omega\right)+4 E_0^2 \left(3 r^5-i r^4 R_4^2 \omega+6 r R_4^4-i R_4^6 \omega\right)\right)-9 r \omega^4 \left(3 E_0^2 r^4 f(r)^3-r^3 f(r)^2  \right.\right.\nonumber\\
    &&\left.\left. \left(3 E_0^2 r+2 i E_0^2 R_4^2 \omega+12 r R_4^4\right)+3 f(r) \left(2 R_4^4 \left(E_0^2+2 r^4\right)+E_0^2 r^4-i E_0^2 r^3 R_4^2 \omega\right) \right.\right.\nonumber\\
    &&\left.\left. -3 E_0^2 \left(r^4+2 R_4^4\right)\right)\right]/\left\{r^3 R_4^6 \left(k^2 (f(r)+2)-3 \omega^2\right){}^3 \left(3 r \omega \left(4 r f(r)-2 r-i R_4^2 \omega\right) \right.\right.\nonumber\\
    &&\left.\left. -k^2 R_4^2 \left(3 i r f(r)-2 i r+R_4^2 \omega\right)\right)\right\},
\end{eqnarray}

\begin{eqnarray}
    Q_1 & = & i \sin^3\theta(r)\left[\left.4 k^2 r \left(132 r+7 i R_4^2 \omega\right)+3 k^4 R_4^4-504 r^2 \omega^2\right)+E_0^2 R_4^2 \left(2 k^2-3 \omega^2\right)^2 \right.\nonumber\\
    && \left(2 r+i R_4^2 \omega\right) \left(2 k^4 \left(5 r^4 R_4^2+2 R_4^6\right)+3 k^2 \omega \left(10 i r^5+3 r^4 R_4^2 \omega+4 i r R_4^4+6 R_4^6 \omega\right) \right.\nonumber\\
    &&\left. +27 i r \omega^3 \left(r^4+2 R_4^4\right)\right)-3 k^2 r^4 f(r)^5 \left(2 k^4 r \left(8 R_4^4 \left(132 r^2-E_0^2 \omega^2\right)+312 E_0^2 r^2+147 i \text{E0}^2 r R_4^2 \omega \right.\right.\nonumber\\
    &&\left.\left.\left. +56 i r R_4^6 \omega\right)+12 k^2 r^2 \omega^2 \left(2 E_0^2 r+19 i E_0^2 R_4^2 \omega-168 r R_4^4\right)+k^6 \left(16 E_0^2 r R_4^4+i E_0^2 R_4^6 \omega+12 r R_4^8\right) \right.\right.\nonumber\\
    &&\left. -576 E_0^2 r^3 \omega^4\right)+f(r)^3 \left(k^8 R_4^4 \left(R_4^4 \left(96 r^5-176 E_0^2 r\right)+208 E_0^2 r^5-2 i R_4^6 \omega \left(11 E_0^2-36 r^4\right) \right.\right.\nonumber\\
    &&\left.\left. +161 i E_0^2 r^4 R_4^2 \omega\right)-54 k^2 r^3 \omega^4 \left(32 E_0^2 r^4+53 i E_0^2 r^3 R_4^2 \omega-4 R_4^4 \left(5 E_0^2 r^2 \omega^2+16 E_0^2-32 r^4\right) \right.\right.\nonumber\\
    &&\left. +152 i r^3 R_4^6 \omega\right)+9 k^4 r^2 \omega^2 \left(152 E_0^2 r^5+380 i E_0^2 r^4 R_4^2 \omega-i R_4^6 \omega \left(25 E_0^2 r^2 \omega^2+184 E_0^2 \right.\right.\nonumber\\
    &&\left.\left.\left. -1176 r^4\right)-16 r R_4^4 \left(3 E_0^2 r^2 \omega^2+E_0^2+4 r^4\right)+56 r^3 R_4^8 \omega^2\right)-6 k^6 r \left(312 E_0^2 r^6+123 i E_0^2 r^5 R_4^2 \omega \right.\right.\nonumber\\
    &&\left. -16 R_4^8 \omega^2 \left(3 E_0^2-7 r^4\right)+2 i r R_4^6 \omega \left(7 E_0^2 r^2 \omega^2-E_0^2+348 r^4\right)+48 R_4^4 \left(2 E_0^2 r^4 \omega^2+13 E_0^2 r^2 \right.\right.\nonumber\\
    &&\left.\left.\left.\left. +4 r^6\right)+18 i r^3 R_4^{10} \omega^3\right)+324 r^6 \omega^6 \left(6 E_0^2 r-7 i E_0^2 R_4^2 \omega+24 r R_4^4\right)\right)+f(r) \left(2 k^2-3 \omega^2\right) \right.\nonumber\\
    &&  \left(E_0^2 \left(k^6 \left(224 r^5 R_4^4+30 i r^4 R_4^6 \omega+224 r R_4^8-36 i R_4^{10} \omega\right)+3 i k^4 R_4^2 \omega \left(28 r^6+144 i r^5 R_4^2 \omega \right.\right.\right.\nonumber\\
    &&\left.\left. +r^2 R_4^4 \left(r^2 \omega^2-136\right)+80 i r R_4^6 \omega+34 R_4^8 \omega^2\right)+18 k^2 r \omega^2 \left(40 r^6+13 i r^5 R_4^2 \omega \right.\right.\nonumber\\
    &&\left.\left.\left.\left. +4 r^2 R_4^4 \left(r^2 \omega^2+20\right)+42 i r R_4^6 \omega-8 R_4^8 \omega^2\right)-324 r^2 \omega^4 \left(r^4+2 R_4^4\right) \left(2 r-i R_4^2 \omega\right)\right) \right.\right.\nonumber\\
    &&\left. +12 r^4 R_4^6 \left(2 k^2-3 \omega^2\right){}^2 \left(2 r+i R_4^2 \omega\right) \left(k^2 R_4^2+3 i r \omega\right)\right)+f(r)^4 \left(E_0^2 r \left(108 k^2 r^5 \omega^4  \right.\right.\nonumber\\
    &&\left. \left(16 r+35 i R_4^2 \omega\right)+k^8 \left(-7 r^4 R_4^4+30 i r^3 R_4^6 \omega+2 R_4^8\right)+3 k^6 r \left(624 r^5-152 i r^4 R_4^2 \omega \right.\right.\nonumber\\
    &&\left.\left. +i R_4^6 \omega \left(17 r^2 \omega^2+152\right)+32 r R_4^4 \left(r^2 \omega^2+33\right)\right)-18 k^4 r^2 \omega^2 \left(76 r^4+7 i r^3 R_4^2 \omega \right.\right.\nonumber\\
    &&\left.\left.\left. +3 R_4^4 \left(9 r^2 \omega^2+56\right)\right)-1944 r^6 \omega^6\right)+12 k^2 r^4 R_4^4 \left(2 k^4 r \left(312 r^2+15 i r R_4^2 \omega-4 R_4^4 \omega^2\right) \right.\right.\nonumber\\
    &&\left.\left. +12 k^2 r^2 \omega^2 \left(2 r+7 i R_4^2 \omega\right)+k^6 \left(16 r R_4^4+i R_4^6 \omega\right)-576 r^3 \omega^4\right)\right)+3 f(r)^2 \left(E_0^2 \left(k^8 \left(40 r^5 R_4^4 \right.\right.\right.\nonumber\\
    &&\left.\left. +66 i r^4 R_4^6 \omega-48 r R_4^8-28 i R_4^{10} \omega\right)+9 k^2 r^2 \omega_1^4 \left(64 r^5-136 i r^4 R_4^2 \omega_1+i R_4^6 \omega_1 \left(11 r^2 \omega^2+184\right) \right.\right.\nonumber\\
    &&\left.\left. +32 r R_4^4 \left(3 r^2 \omega^2+4\right)\right)+k^6 \left(96 r^7+204 i r^6 R_4^2 \omega-i r^2 R_4^6 \omega \left(131 r^2 \omega^2-408\right) \right.\right.\nonumber\\
    &&\left.\left. +8 r^3 R_4^4 \left(13 r^2 \omega^2+24\right)+288 r R_4^8 \omega^2+50 i R_4^{10} \omega^3\right)+6 k^4 r \omega^2 \left(8 r^6+33 i r^5 R_4^2 \omega \right.\right.\nonumber\\
    &&\left.\left. -i r R_4^6 \omega \left(r^2 \omega^2+230\right)+R_4^4 \left(16 r^2-119 r^4 \omega^2\right)-54 R_4^8 \omega^2\right)-27 r^3 \omega^6 \left(24 r^4-46 i r^3 R_4^2 \omega \right.\right.\nonumber\\
    &&\left.\left.\left. +R_4^4 \left(11 r^2 \omega^2+48\right)\right)\right)-4 r^4 R_4^4 \left(2 k^2-3 \omega^2\right) \left(i k^4 R_4^2 \omega \left(-68 r^2+32 i r R_4^2 \omega+9 R_4^4 \omega^2\right) \right.\right.\nonumber\\
    &&\left.\left.\left. +30 k^2 r^2 \omega^2 \left(8 r+i R_4^2 \omega\right)+k^6 \left(32 r R_4^4-6 i R_4^6 \omega\right)+108 r^2 \omega^4 \left(-2 r+i R_4^2 \omega\right)\right)\right)\right]/ \nonumber\\
    && \left\{12 r^6 R_4^6 f(r)^2 \left(k^2 (f(r)+2)-3 \omega^2\right){}^3 \left(3 r \omega \left(4 r f(r)-2 r-i R_4^2 \omega\right)-k^2 R_4^2 \left(3 i r f(r) \right.\right.\right.\nonumber\\
    &&\left.\left.\left. -2 i r+R_4^2 \omega\right)\right)\right\}.
\end{eqnarray}


\end{subequations}

\section{Coefficients of Master Equations: shear mode}\label{app_shear}
The coefficients of master equation \eqref{shear_eq} can be expressed as,
\begin{subequations}
\begin{eqnarray}
    \mathcal{P}_{sh}(r,\omega,k) & = & P_0(r,\omega,k)+n_f P_1(r,\omega,k),\\
     \mathcal{Q}_{sh}(r,\omega,k) & = & Q_0(r,\omega,k)+n_f Q_1(r,\omega,k),
\end{eqnarray}
where these $P$s and $Q$s are given below.
\begin{eqnarray}
        P_0 & = & \frac{-\omega  f(r) \left(5 r \omega +2 i k^2 R_4^2\right)+9 k^2 r f(r)^2-4 r \omega ^2+2 i R_4^2 \omega ^3}{r^2 f(r) \left(k^2 f(r)-\omega ^2\right)}, \\
        Q_0 & = & \left[12 k^2 r^2 f(r)^2-f(r) \left(k^4 R_4^4+11 i k^2 r R_4^2 \omega +4 r^2 \omega ^2\right)+\omega  \left(i r R_4^2 \left(4 k^2+7 \omega ^2\right)+k^2 R_4^4 \omega  \right.\right.\nonumber\\
        &&\left.\left.  -8 r^2 \omega \right)\right]/\left\{r^4 f(r) \left(k^2 f(r)-\omega ^2\right)\right\},
\end{eqnarray}

\begin{eqnarray}
        P_1 & = & -\left[i k^2 \omega^2 (f(r)-1) \sin ^3(\Theta (r)) \left(2 R_4^4 \left(E_0^2-2 r^4 f(r)\right)+E_0^2 r^4 \left(f(r)^2+1\right)\right)^2\right]/\left\{r^3 R_4^6 \right.\nonumber\\
        &&  \left(\omega^2-k^2 f(r)\right)^2 \left(-r \omega \left(2 R_4^4 \left(5 f(r) \left(E_0^2-2 r^4 f(r)\right)-4 E_0^2\right)+E_0^2 r^4 (f(r) (f(r) (f(r)+4)+9) \right.\right.\nonumber\\
        && \left.\left.\left. -4)\right)+i k^2 R_4^2 f(r) \left(2 R_4^4 \left(E_0^2-2 r^4 f(r)\right)+E_0^2 r^4 \left(f(r)^2+1\right)\right)\right)\right\},
\end{eqnarray}
    
\begin{eqnarray}
         Q_1 & = & \left[i \sin ^3(\theta (r)) \left(2 R_4^4 \left(E_0^2-2 r^4 f(r)\right)+E_0^2 r^4 \left(f(r)^2+1\right)\right) \left(i k^4 r R_4^2 \omega f(r)^2 \left(2 R_4^4 \left(f(r) \left(9 E_0^2 \right.\right.\right.\right.\right.\nonumber\\
         &&\left.\left.\left.\left.\left. -18 r^4 f(r)+8 r^4\right)-8 E_0^2\right)+E_0^2 r^4 \left(5 f(r)^3+13 f(r)-8\right)\right)+k^2 R_4^4 \omega^4 f(r) \left(2 R_4^4 \left(E_0^2 \right.\right.\right.\right.\nonumber\\
         &&\left.\left.\left.\left. -2 r^4 f(r)\right)+E_0^2 r^4 \left(f(r)^2+1\right)\right)-2 i k^2 r R_4^2 \omega^3 f(r) \left(2 R_4^4 \left(f(r) \left(9 E_0^2-18 r^4 f(r)+8 r^4\right) \right.\right.\right.\right.\nonumber\\
         &&\left.\left.\left.\left. -8 E_0^2\right)+E_0^2 r^4 \left(5 f(r)^3+13 f(r)-8\right)\right)+k^6 R_4^4 f(r)^3 \left(2 R_4^4 \left(E_0^2-2 r^4 f(r)\right)+E_0^2 r^4 \left(f(r)^2 \right.\right.\right.\right.\nonumber\\
         &&\left.\left.\left.\left. +1\right)\right)+2 k^2 \omega^2 f(r)^2 \left(2 R_4^4 \left(E_0^2-2 r^4 f(r)\right)+E_0^2 r^4 \left(f(r)^2+1\right)\right) \left(4 r^2 (f(r)-1)-k^2 R_4^4\right) \right.\right.\nonumber\\
         &&\left.\left. +i r R_4^2 \omega^5 \left(2 R_4^4 \left(5 f(r) \left(E_0^2-2 r^4 f(r)\right)-4 E_0^2\right)+E_0^2 r^4 (f(r) (f(r) (f(r)+4)+9)-4)\right)\right)\right] \nonumber\\
         && /\left\{4 r^6 R_4^6 f(r)^2 \left(\omega^2-k^2 f(r)\right)^2 \left(r \omega \left(2 R_4^4 \left(5 f(r) \left(E_0^2-2 r^4 f(r)\right)-4 E_0^2\right)+E_0^2 r^4 (f(r) (f(r) \right.\right.\right.\nonumber\\
         &&\left.\left.\left. (f(r)+4)+9)-4)\right)-i k^2 R_4^2 f(r) \left(2 R_4^4 \left(E_0^2-2 r^4 f(r)\right)+E_0^2 r^4 \left(f(r)^2+1\right)\right)\right)\right\}.
\end{eqnarray}
\end{subequations}

\section{Coefficients of Master Equations: tensor mode}\label{app_tensor}
The coefficients of master equation \eqref{shear_eq} can be expressed as,
\begin{subequations}
\begin{eqnarray}
    \mathcal{P}_{ten} & = & \frac{rf(r)+4 r-2 i R_4^2 \omega }{r^2 f(r)},\\
    \mathcal{Q}_{ten} & = & -\frac{k^2 R_4^4+3 i r R_4^2 \omega }{r^4 f(r)}+n_f\left[\frac{\sin^3\theta(r)}{r^2 f(r)} \right. \nonumber\\ 
    &&\left. -\frac{r^5 f(r)^2+\left(r-r_h \sin^3\theta_0\right) \left(r^4+r^3 r_h \sin^3\theta_0+2 R_4^4\right)}{4r^7 R_4^4 f(r)^2}E_0^2\sin^3\theta(r)\right].
\end{eqnarray}
\end{subequations}
\end{appendix}

\twocolumngrid
\bibliographystyle{apsrev4-1}
\bibliography{prd_chaosbrane}

\begin{thebibliography}{53}%
\makeatletter
\providecommand \@ifxundefined [1]{%
 \@ifx{#1\undefined}
}%
\providecommand \@ifnum [1]{%
 \ifnum #1\expandafter \@firstoftwo
 \else \expandafter \@secondoftwo
 \fi
}%
\providecommand \@ifx [1]{%
 \ifx #1\expandafter \@firstoftwo
 \else \expandafter \@secondoftwo
 \fi
}%
\providecommand \natexlab [1]{#1}%
\providecommand \enquote  [1]{``#1''}%
\providecommand \bibnamefont  [1]{#1}%
\providecommand \bibfnamefont [1]{#1}%
\providecommand \citenamefont [1]{#1}%
\providecommand \href@noop [0]{\@secondoftwo}%
\providecommand \href [0]{\begingroup \@sanitize@url \@href}%
\providecommand \@href[1]{\@@startlink{#1}\@@href}%
\providecommand \@@href[1]{\endgroup#1\@@endlink}%
\providecommand \@sanitize@url [0]{\catcode `\\12\catcode `\$12\catcode
  `\&12\catcode `\#12\catcode `\^12\catcode `\_12\catcode `\%12\relax}%
\providecommand \@@startlink[1]{}%
\providecommand \@@endlink[0]{}%
\providecommand \url  [0]{\begingroup\@sanitize@url \@url }%
\providecommand \@url [1]{\endgroup\@href {#1}{\urlprefix }}%
\providecommand \urlprefix  [0]{URL }%
\providecommand \Eprint [0]{\href }%
\providecommand \doibase [0]{http://dx.doi.org/}%
\providecommand \selectlanguage [0]{\@gobble}%
\providecommand \bibinfo  [0]{\@secondoftwo}%
\providecommand \bibfield  [0]{\@secondoftwo}%
\providecommand \translation [1]{[#1]}%
\providecommand \BibitemOpen [0]{}%
\providecommand \bibitemStop [0]{}%
\providecommand \bibitemNoStop [0]{.\EOS\space}%
\providecommand \EOS [0]{\spacefactor3000\relax}%
\providecommand \BibitemShut  [1]{\csname bibitem#1\endcsname}%
\let\auto@bib@innerbib\@empty
\bibitem [{\citenamefont {Maldacena}(1998)}]{Maldacena:1997re}%
  \BibitemOpen
  \bibfield  {author} {\bibinfo {author} {\bibfnamefont {J.~M.}\ \bibnamefont
  {Maldacena}},\ }\href {\doibase 10.4310/ATMP.1998.v2.n2.a1} {\bibfield
  {journal} {\bibinfo  {journal} {Adv. Theor. Math. Phys.}\ }\textbf {\bibinfo
  {volume} {2}},\ \bibinfo {pages} {231} (\bibinfo {year} {1998})},\ \Eprint
  {http://arxiv.org/abs/hep-th/9711200} {arXiv:hep-th/9711200} \BibitemShut
  {NoStop}%
\bibitem [{\citenamefont {Maldacena}(2003)}]{Maldacena:2001kr}%
  \BibitemOpen
  \bibfield  {author} {\bibinfo {author} {\bibfnamefont {J.~M.}\ \bibnamefont
  {Maldacena}},\ }\href {\doibase 10.1088/1126-6708/2003/04/021} {\bibfield
  {journal} {\bibinfo  {journal} {JHEP}\ }\textbf {\bibinfo {volume} {04}},\
  \bibinfo {pages} {021} (\bibinfo {year} {2003})},\ \Eprint
  {http://arxiv.org/abs/hep-th/0106112} {arXiv:hep-th/0106112} \BibitemShut
  {NoStop}%
\bibitem [{\citenamefont {Shenker}\ and\ \citenamefont
  {Stanford}(2014{\natexlab{a}})}]{Shenker:2013pqa}%
  \BibitemOpen
  \bibfield  {author} {\bibinfo {author} {\bibfnamefont {S.~H.}\ \bibnamefont
  {Shenker}}\ and\ \bibinfo {author} {\bibfnamefont {D.}~\bibnamefont
  {Stanford}},\ }\href {\doibase 10.1007/JHEP03(2014)067} {\bibfield  {journal}
  {\bibinfo  {journal} {JHEP}\ }\textbf {\bibinfo {volume} {03}},\ \bibinfo
  {pages} {067} (\bibinfo {year} {2014}{\natexlab{a}})},\ \Eprint
  {http://arxiv.org/abs/1306.0622} {arXiv:1306.0622 [hep-th]} \BibitemShut
  {NoStop}%
\bibitem [{\citenamefont {Shenker}\ and\ \citenamefont
  {Stanford}(2014{\natexlab{b}})}]{Shenker:2013yza}%
  \BibitemOpen
  \bibfield  {author} {\bibinfo {author} {\bibfnamefont {S.~H.}\ \bibnamefont
  {Shenker}}\ and\ \bibinfo {author} {\bibfnamefont {D.}~\bibnamefont
  {Stanford}},\ }\href {\doibase 10.1007/JHEP12(2014)046} {\bibfield  {journal}
  {\bibinfo  {journal} {JHEP}\ }\textbf {\bibinfo {volume} {12}},\ \bibinfo
  {pages} {046} (\bibinfo {year} {2014}{\natexlab{b}})},\ \Eprint
  {http://arxiv.org/abs/1312.3296} {arXiv:1312.3296 [hep-th]} \BibitemShut
  {NoStop}%
\bibitem [{\citenamefont {Shenker}\ and\ \citenamefont
  {Stanford}(2015)}]{Shenker:2014cwa}%
  \BibitemOpen
  \bibfield  {author} {\bibinfo {author} {\bibfnamefont {S.~H.}\ \bibnamefont
  {Shenker}}\ and\ \bibinfo {author} {\bibfnamefont {D.}~\bibnamefont
  {Stanford}},\ }\href {\doibase 10.1007/JHEP05(2015)132} {\bibfield  {journal}
  {\bibinfo  {journal} {JHEP}\ }\textbf {\bibinfo {volume} {05}},\ \bibinfo
  {pages} {132} (\bibinfo {year} {2015})},\ \Eprint
  {http://arxiv.org/abs/1412.6087} {arXiv:1412.6087 [hep-th]} \BibitemShut
  {NoStop}%
\bibitem [{\citenamefont {Roberts}\ \emph {et~al.}(2015)\citenamefont
  {Roberts}, \citenamefont {Stanford},\ and\ \citenamefont
  {Susskind}}]{Roberts:2014isa}%
  \BibitemOpen
  \bibfield  {author} {\bibinfo {author} {\bibfnamefont {D.~A.}\ \bibnamefont
  {Roberts}}, \bibinfo {author} {\bibfnamefont {D.}~\bibnamefont {Stanford}}, \
  and\ \bibinfo {author} {\bibfnamefont {L.}~\bibnamefont {Susskind}},\ }\href
  {\doibase 10.1007/JHEP03(2015)051} {\bibfield  {journal} {\bibinfo  {journal}
  {JHEP}\ }\textbf {\bibinfo {volume} {03}},\ \bibinfo {pages} {051} (\bibinfo
  {year} {2015})},\ \Eprint {http://arxiv.org/abs/1409.8180} {arXiv:1409.8180
  [hep-th]} \BibitemShut {NoStop}%
\bibitem [{\citenamefont {Blake}\ \emph
  {et~al.}(2018{\natexlab{a}})\citenamefont {Blake}, \citenamefont {Lee},\ and\
  \citenamefont {Liu}}]{Blake:2017ris}%
  \BibitemOpen
  \bibfield  {author} {\bibinfo {author} {\bibfnamefont {M.}~\bibnamefont
  {Blake}}, \bibinfo {author} {\bibfnamefont {H.}~\bibnamefont {Lee}}, \ and\
  \bibinfo {author} {\bibfnamefont {H.}~\bibnamefont {Liu}},\ }\href {\doibase
  10.1007/JHEP10(2018)127} {\bibfield  {journal} {\bibinfo  {journal} {JHEP}\
  }\textbf {\bibinfo {volume} {10}},\ \bibinfo {pages} {127} (\bibinfo {year}
  {2018}{\natexlab{a}})},\ \Eprint {http://arxiv.org/abs/1801.00010}
  {arXiv:1801.00010 [hep-th]} \BibitemShut {NoStop}%
\bibitem [{\citenamefont {Blake}\ \emph
  {et~al.}(2018{\natexlab{b}})\citenamefont {Blake}, \citenamefont {Davison},
  \citenamefont {Grozdanov},\ and\ \citenamefont {Liu}}]{Blake:2018leo}%
  \BibitemOpen
  \bibfield  {author} {\bibinfo {author} {\bibfnamefont {M.}~\bibnamefont
  {Blake}}, \bibinfo {author} {\bibfnamefont {R.~A.}\ \bibnamefont {Davison}},
  \bibinfo {author} {\bibfnamefont {S.}~\bibnamefont {Grozdanov}}, \ and\
  \bibinfo {author} {\bibfnamefont {H.}~\bibnamefont {Liu}},\ }\href {\doibase
  10.1007/JHEP10(2018)035} {\bibfield  {journal} {\bibinfo  {journal} {JHEP}\
  }\textbf {\bibinfo {volume} {10}},\ \bibinfo {pages} {035} (\bibinfo {year}
  {2018}{\natexlab{b}})},\ \Eprint {http://arxiv.org/abs/1809.01169}
  {arXiv:1809.01169 [hep-th]} \BibitemShut {NoStop}%
\bibitem [{\citenamefont {Grozdanov}(2019)}]{Grozdanov:2018kkt}%
  \BibitemOpen
  \bibfield  {author} {\bibinfo {author} {\bibfnamefont {S.}~\bibnamefont
  {Grozdanov}},\ }\href {\doibase 10.1007/JHEP01(2019)048} {\bibfield
  {journal} {\bibinfo  {journal} {JHEP}\ }\textbf {\bibinfo {volume} {01}},\
  \bibinfo {pages} {048} (\bibinfo {year} {2019})},\ \Eprint
  {http://arxiv.org/abs/1811.09641} {arXiv:1811.09641 [hep-th]} \BibitemShut
  {NoStop}%
\bibitem [{\citenamefont {Ahn}\ \emph {et~al.}(2019)\citenamefont {Ahn},
  \citenamefont {Jahnke}, \citenamefont {Jeong},\ and\ \citenamefont
  {Kim}}]{Ahn:2019rnq}%
  \BibitemOpen
  \bibfield  {author} {\bibinfo {author} {\bibfnamefont {Y.}~\bibnamefont
  {Ahn}}, \bibinfo {author} {\bibfnamefont {V.}~\bibnamefont {Jahnke}},
  \bibinfo {author} {\bibfnamefont {H.-S.}\ \bibnamefont {Jeong}}, \ and\
  \bibinfo {author} {\bibfnamefont {K.-Y.}\ \bibnamefont {Kim}},\ }\href
  {\doibase 10.1007/JHEP10(2019)257} {\bibfield  {journal} {\bibinfo  {journal}
  {JHEP}\ }\textbf {\bibinfo {volume} {10}},\ \bibinfo {pages} {257} (\bibinfo
  {year} {2019})},\ \Eprint {http://arxiv.org/abs/1907.08030} {arXiv:1907.08030
  [hep-th]} \BibitemShut {NoStop}%
\bibitem [{\citenamefont {Blake}\ \emph {et~al.}(2020)\citenamefont {Blake},
  \citenamefont {Davison},\ and\ \citenamefont {Vegh}}]{Blake:2019otz}%
  \BibitemOpen
  \bibfield  {author} {\bibinfo {author} {\bibfnamefont {M.}~\bibnamefont
  {Blake}}, \bibinfo {author} {\bibfnamefont {R.~A.}\ \bibnamefont {Davison}},
  \ and\ \bibinfo {author} {\bibfnamefont {D.}~\bibnamefont {Vegh}},\ }\href
  {\doibase 10.1007/JHEP01(2020)077} {\bibfield  {journal} {\bibinfo  {journal}
  {JHEP}\ }\textbf {\bibinfo {volume} {01}},\ \bibinfo {pages} {077} (\bibinfo
  {year} {2020})},\ \Eprint {http://arxiv.org/abs/1904.12883} {arXiv:1904.12883
  [hep-th]} \BibitemShut {NoStop}%
\bibitem [{\citenamefont {Natsuume}\ and\ \citenamefont
  {Okamura}(2019{\natexlab{a}})}]{Natsuume:2019xcy}%
  \BibitemOpen
  \bibfield  {author} {\bibinfo {author} {\bibfnamefont {M.}~\bibnamefont
  {Natsuume}}\ and\ \bibinfo {author} {\bibfnamefont {T.}~\bibnamefont
  {Okamura}},\ }\href {\doibase 10.1007/JHEP12(2019)139} {\bibfield  {journal}
  {\bibinfo  {journal} {JHEP}\ }\textbf {\bibinfo {volume} {12}},\ \bibinfo
  {pages} {139} (\bibinfo {year} {2019}{\natexlab{a}})},\ \Eprint
  {http://arxiv.org/abs/1905.12015} {arXiv:1905.12015 [hep-th]} \BibitemShut
  {NoStop}%
\bibitem [{\citenamefont {Blake}\ and\ \citenamefont
  {Davison}(2022)}]{Blake:2021hjj}%
  \BibitemOpen
  \bibfield  {author} {\bibinfo {author} {\bibfnamefont {M.}~\bibnamefont
  {Blake}}\ and\ \bibinfo {author} {\bibfnamefont {R.~A.}\ \bibnamefont
  {Davison}},\ }\href {\doibase 10.1007/JHEP01(2022)013} {\bibfield  {journal}
  {\bibinfo  {journal} {JHEP}\ }\textbf {\bibinfo {volume} {01}},\ \bibinfo
  {pages} {013} (\bibinfo {year} {2022})},\ \Eprint
  {http://arxiv.org/abs/2111.11093} {arXiv:2111.11093 [hep-th]} \BibitemShut
  {NoStop}%
\bibitem [{\citenamefont {Wu}(2019)}]{Wu:2019esr}%
  \BibitemOpen
  \bibfield  {author} {\bibinfo {author} {\bibfnamefont {X.}~\bibnamefont
  {Wu}},\ }\href {\doibase 10.1007/JHEP12(2019)140} {\bibfield  {journal}
  {\bibinfo  {journal} {JHEP}\ }\textbf {\bibinfo {volume} {12}},\ \bibinfo
  {pages} {140} (\bibinfo {year} {2019})},\ \Eprint
  {http://arxiv.org/abs/1909.10223} {arXiv:1909.10223 [hep-th]} \BibitemShut
  {NoStop}%
\bibitem [{\citenamefont {Sil}(2021)}]{Sil:2020jhr}%
  \BibitemOpen
  \bibfield  {author} {\bibinfo {author} {\bibfnamefont {K.}~\bibnamefont
  {Sil}},\ }\href {\doibase 10.1007/JHEP03(2021)232} {\bibfield  {journal}
  {\bibinfo  {journal} {JHEP}\ }\textbf {\bibinfo {volume} {03}},\ \bibinfo
  {pages} {232} (\bibinfo {year} {2021})},\ \Eprint
  {http://arxiv.org/abs/2012.07710} {arXiv:2012.07710 [hep-th]} \BibitemShut
  {NoStop}%
\bibitem [{\citenamefont {Li}\ \emph {et~al.}(2019)\citenamefont {Li},
  \citenamefont {Lin},\ and\ \citenamefont {Mei}}]{Li:2019bgc}%
  \BibitemOpen
  \bibfield  {author} {\bibinfo {author} {\bibfnamefont {W.}~\bibnamefont
  {Li}}, \bibinfo {author} {\bibfnamefont {S.}~\bibnamefont {Lin}}, \ and\
  \bibinfo {author} {\bibfnamefont {J.}~\bibnamefont {Mei}},\ }\href {\doibase
  10.1103/PhysRevD.100.046012} {\bibfield  {journal} {\bibinfo  {journal}
  {Phys. Rev. D}\ }\textbf {\bibinfo {volume} {100}},\ \bibinfo {pages}
  {046012} (\bibinfo {year} {2019})},\ \Eprint
  {http://arxiv.org/abs/1905.07684} {arXiv:1905.07684 [hep-th]} \BibitemShut
  {NoStop}%
\bibitem [{\citenamefont {Amano}\ \emph {et~al.}(2023)\citenamefont {Amano},
  \citenamefont {Blake}, \citenamefont {Cartwright}, \citenamefont {Kaminski},\
  and\ \citenamefont {Thompson}}]{Amano:2022mlu}%
  \BibitemOpen
  \bibfield  {author} {\bibinfo {author} {\bibfnamefont {M.~A.~G.}\
  \bibnamefont {Amano}}, \bibinfo {author} {\bibfnamefont {M.}~\bibnamefont
  {Blake}}, \bibinfo {author} {\bibfnamefont {C.}~\bibnamefont {Cartwright}},
  \bibinfo {author} {\bibfnamefont {M.}~\bibnamefont {Kaminski}}, \ and\
  \bibinfo {author} {\bibfnamefont {A.~P.}\ \bibnamefont {Thompson}},\ }\href
  {\doibase 10.1007/JHEP02(2023)253} {\bibfield  {journal} {\bibinfo  {journal}
  {JHEP}\ }\textbf {\bibinfo {volume} {02}},\ \bibinfo {pages} {253} (\bibinfo
  {year} {2023})},\ \Eprint {http://arxiv.org/abs/2211.00016} {arXiv:2211.00016
  [hep-th]} \BibitemShut {NoStop}%
\bibitem [{\citenamefont {Ramirez}(2021)}]{Ramirez:2020qer}%
  \BibitemOpen
  \bibfield  {author} {\bibinfo {author} {\bibfnamefont {D.~M.}\ \bibnamefont
  {Ramirez}},\ }\href {\doibase 10.1007/JHEP12(2021)006} {\bibfield  {journal}
  {\bibinfo  {journal} {JHEP}\ }\textbf {\bibinfo {volume} {12}},\ \bibinfo
  {pages} {006} (\bibinfo {year} {2021})},\ \Eprint
  {http://arxiv.org/abs/2009.00500} {arXiv:2009.00500 [hep-th]} \BibitemShut
  {NoStop}%
\bibitem [{\citenamefont {Haehl}\ \emph {et~al.}(2019)\citenamefont {Haehl},
  \citenamefont {Reeves},\ and\ \citenamefont {Rozali}}]{Haehl:2019eae}%
  \BibitemOpen
  \bibfield  {author} {\bibinfo {author} {\bibfnamefont {F.~M.}\ \bibnamefont
  {Haehl}}, \bibinfo {author} {\bibfnamefont {W.}~\bibnamefont {Reeves}}, \
  and\ \bibinfo {author} {\bibfnamefont {M.}~\bibnamefont {Rozali}},\ }\href
  {\doibase 10.1007/JHEP11(2019)102} {\bibfield  {journal} {\bibinfo  {journal}
  {JHEP}\ }\textbf {\bibinfo {volume} {11}},\ \bibinfo {pages} {102} (\bibinfo
  {year} {2019})},\ \Eprint {http://arxiv.org/abs/1909.05847} {arXiv:1909.05847
  [hep-th]} \BibitemShut {NoStop}%
\bibitem [{\citenamefont {Das}\ \emph {et~al.}(2019)\citenamefont {Das},
  \citenamefont {Ezhuthachan},\ and\ \citenamefont {Kundu}}]{Das:2019tga}%
  \BibitemOpen
  \bibfield  {author} {\bibinfo {author} {\bibfnamefont {S.}~\bibnamefont
  {Das}}, \bibinfo {author} {\bibfnamefont {B.}~\bibnamefont {Ezhuthachan}}, \
  and\ \bibinfo {author} {\bibfnamefont {A.}~\bibnamefont {Kundu}},\ }\href
  {\doibase 10.1007/JHEP12(2019)141} {\bibfield  {journal} {\bibinfo  {journal}
  {JHEP}\ }\textbf {\bibinfo {volume} {12}},\ \bibinfo {pages} {141} (\bibinfo
  {year} {2019})},\ \Eprint {http://arxiv.org/abs/1907.08763} {arXiv:1907.08763
  [hep-th]} \BibitemShut {NoStop}%
\bibitem [{\citenamefont {Liu}\ and\ \citenamefont {Raju}(2020)}]{Liu:2020yaf}%
  \BibitemOpen
  \bibfield  {author} {\bibinfo {author} {\bibfnamefont {Y.}~\bibnamefont
  {Liu}}\ and\ \bibinfo {author} {\bibfnamefont {A.}~\bibnamefont {Raju}},\
  }\href {\doibase 10.1007/JHEP12(2020)027} {\bibfield  {journal} {\bibinfo
  {journal} {JHEP}\ }\textbf {\bibinfo {volume} {12}},\ \bibinfo {pages} {027}
  (\bibinfo {year} {2020})},\ \Eprint {http://arxiv.org/abs/2005.08508}
  {arXiv:2005.08508 [hep-th]} \BibitemShut {NoStop}%
\bibitem [{\citenamefont {Abbasi}\ and\ \citenamefont
  {Tabatabaei}(2020)}]{Abbasi:2019rhy}%
  \BibitemOpen
  \bibfield  {author} {\bibinfo {author} {\bibfnamefont {N.}~\bibnamefont
  {Abbasi}}\ and\ \bibinfo {author} {\bibfnamefont {J.}~\bibnamefont
  {Tabatabaei}},\ }\href {\doibase 10.1007/JHEP03(2020)050} {\bibfield
  {journal} {\bibinfo  {journal} {JHEP}\ }\textbf {\bibinfo {volume} {03}},\
  \bibinfo {pages} {050} (\bibinfo {year} {2020})},\ \Eprint
  {http://arxiv.org/abs/1910.13696} {arXiv:1910.13696 [hep-th]} \BibitemShut
  {NoStop}%
\bibitem [{\citenamefont {Mahish}\ and\ \citenamefont
  {Sil}(2022)}]{Mahish:2022xjz}%
  \BibitemOpen
  \bibfield  {author} {\bibinfo {author} {\bibfnamefont {S.}~\bibnamefont
  {Mahish}}\ and\ \bibinfo {author} {\bibfnamefont {K.}~\bibnamefont {Sil}},\
  }\href {\doibase 10.1007/JHEP08(2022)041} {\bibfield  {journal} {\bibinfo
  {journal} {JHEP}\ }\textbf {\bibinfo {volume} {08}},\ \bibinfo {pages} {041}
  (\bibinfo {year} {2022})},\ \Eprint {http://arxiv.org/abs/2202.05865}
  {arXiv:2202.05865 [hep-th]} \BibitemShut {NoStop}%
\bibitem [{\citenamefont {Natsuume}\ and\ \citenamefont
  {Okamura}(2019{\natexlab{b}})}]{Natsuume:2019vcv}%
  \BibitemOpen
  \bibfield  {author} {\bibinfo {author} {\bibfnamefont {M.}~\bibnamefont
  {Natsuume}}\ and\ \bibinfo {author} {\bibfnamefont {T.}~\bibnamefont
  {Okamura}},\ }\href {\doibase 10.1103/PhysRevD.100.126012} {\bibfield
  {journal} {\bibinfo  {journal} {Phys. Rev. D}\ }\textbf {\bibinfo {volume}
  {100}},\ \bibinfo {pages} {126012} (\bibinfo {year} {2019}{\natexlab{b}})},\
  \Eprint {http://arxiv.org/abs/1909.09168} {arXiv:1909.09168 [hep-th]}
  \BibitemShut {NoStop}%
\bibitem [{\citenamefont {Baishya}\ and\ \citenamefont
  {Nayek}(2024)}]{BAISHYA2024116521}%
  \BibitemOpen
  \bibfield  {author} {\bibinfo {author} {\bibfnamefont {B.}~\bibnamefont
  {Baishya}}\ and\ \bibinfo {author} {\bibfnamefont {K.}~\bibnamefont
  {Nayek}},\ }\href {\doibase https://doi.org/10.1016/j.nuclphysb.2024.116521}
  {\bibfield  {journal} {\bibinfo  {journal} {Nuclear Physics B}\ }\textbf
  {\bibinfo {volume} {1001}},\ \bibinfo {pages} {116521} (\bibinfo {year}
  {2024})}\BibitemShut {NoStop}%
\bibitem [{\citenamefont {Ceplak}\ and\ \citenamefont
  {Vegh}(2021)}]{Ceplak:2021efc}%
  \BibitemOpen
  \bibfield  {author} {\bibinfo {author} {\bibfnamefont {N.}~\bibnamefont
  {Ceplak}}\ and\ \bibinfo {author} {\bibfnamefont {D.}~\bibnamefont {Vegh}},\
  }\href {\doibase 10.1103/PhysRevD.103.106009} {\bibfield  {journal} {\bibinfo
   {journal} {Phys. Rev. D}\ }\textbf {\bibinfo {volume} {103}},\ \bibinfo
  {pages} {106009} (\bibinfo {year} {2021})},\ \Eprint
  {http://arxiv.org/abs/2101.01490} {arXiv:2101.01490 [hep-th]} \BibitemShut
  {NoStop}%
\bibitem [{\citenamefont {Yuan}\ and\ \citenamefont {Ge}(2021)}]{Yuan:2020fvv}%
  \BibitemOpen
  \bibfield  {author} {\bibinfo {author} {\bibfnamefont {H.}~\bibnamefont
  {Yuan}}\ and\ \bibinfo {author} {\bibfnamefont {X.-H.}\ \bibnamefont {Ge}},\
  }\href {\doibase 10.1007/JHEP06(2021)165} {\bibfield  {journal} {\bibinfo
  {journal} {JHEP}\ }\textbf {\bibinfo {volume} {06}},\ \bibinfo {pages} {165}
  (\bibinfo {year} {2021})},\ \Eprint {http://arxiv.org/abs/2012.15396}
  {arXiv:2012.15396 [hep-th]} \BibitemShut {NoStop}%
\bibitem [{\citenamefont {Kim}\ \emph {et~al.}(2021)\citenamefont {Kim},
  \citenamefont {Lee},\ and\ \citenamefont {Nishida}}]{Kim:2020url}%
  \BibitemOpen
  \bibfield  {author} {\bibinfo {author} {\bibfnamefont {K.-Y.}\ \bibnamefont
  {Kim}}, \bibinfo {author} {\bibfnamefont {K.-S.}\ \bibnamefont {Lee}}, \ and\
  \bibinfo {author} {\bibfnamefont {M.}~\bibnamefont {Nishida}},\ }\href
  {\doibase 10.1007/JHEP04(2021)092} {\bibfield  {journal} {\bibinfo  {journal}
  {JHEP}\ }\textbf {\bibinfo {volume} {04}},\ \bibinfo {pages} {092} (\bibinfo
  {year} {2021})},\ \bibinfo {note} {[Erratum: JHEP 04, 229 (2021)]},\ \Eprint
  {http://arxiv.org/abs/2011.13716} {arXiv:2011.13716 [hep-th]} \BibitemShut
  {NoStop}%
\bibitem [{\citenamefont {Grozdanov}(2021)}]{Grozdanov:2020koi}%
  \BibitemOpen
  \bibfield  {author} {\bibinfo {author} {\bibfnamefont {S.}~\bibnamefont
  {Grozdanov}},\ }\href {\doibase 10.1103/PhysRevLett.126.051601} {\bibfield
  {journal} {\bibinfo  {journal} {Phys. Rev. Lett.}\ }\textbf {\bibinfo
  {volume} {126}},\ \bibinfo {pages} {051601} (\bibinfo {year} {2021})},\
  \Eprint {http://arxiv.org/abs/2008.00888} {arXiv:2008.00888 [hep-th]}
  \BibitemShut {NoStop}%
\bibitem [{\citenamefont {Jeong}\ \emph {et~al.}(2021)\citenamefont {Jeong},
  \citenamefont {Kim},\ and\ \citenamefont {Sun}}]{Jeong:2021zhz}%
  \BibitemOpen
  \bibfield  {author} {\bibinfo {author} {\bibfnamefont {H.-S.}\ \bibnamefont
  {Jeong}}, \bibinfo {author} {\bibfnamefont {K.-Y.}\ \bibnamefont {Kim}}, \
  and\ \bibinfo {author} {\bibfnamefont {Y.-W.}\ \bibnamefont {Sun}},\ }\href
  {\doibase 10.1007/JHEP07(2021)105} {\bibfield  {journal} {\bibinfo  {journal}
  {JHEP}\ }\textbf {\bibinfo {volume} {07}},\ \bibinfo {pages} {105} (\bibinfo
  {year} {2021})},\ \Eprint {http://arxiv.org/abs/2104.13084} {arXiv:2104.13084
  [hep-th]} \BibitemShut {NoStop}%
\bibitem [{\citenamefont {Abbasi}\ and\ \citenamefont
  {Kaminski}(2021)}]{Abbasi:2020xli}%
  \BibitemOpen
  \bibfield  {author} {\bibinfo {author} {\bibfnamefont {N.}~\bibnamefont
  {Abbasi}}\ and\ \bibinfo {author} {\bibfnamefont {M.}~\bibnamefont
  {Kaminski}},\ }\href {\doibase 10.1007/JHEP03(2021)265} {\bibfield  {journal}
  {\bibinfo  {journal} {JHEP}\ }\textbf {\bibinfo {volume} {03}},\ \bibinfo
  {pages} {265} (\bibinfo {year} {2021})},\ \Eprint
  {http://arxiv.org/abs/2012.15820} {arXiv:2012.15820 [hep-th]} \BibitemShut
  {NoStop}%
\bibitem [{\citenamefont {Kim}\ \emph {et~al.}(2022)\citenamefont {Kim},
  \citenamefont {Lee},\ and\ \citenamefont {Nishida}}]{Kim:2021xdz}%
  \BibitemOpen
  \bibfield  {author} {\bibinfo {author} {\bibfnamefont {K.-Y.}\ \bibnamefont
  {Kim}}, \bibinfo {author} {\bibfnamefont {K.-S.}\ \bibnamefont {Lee}}, \ and\
  \bibinfo {author} {\bibfnamefont {M.}~\bibnamefont {Nishida}},\ }\href
  {\doibase 10.1103/PhysRevD.105.126011} {\bibfield  {journal} {\bibinfo
  {journal} {Phys. Rev. D}\ }\textbf {\bibinfo {volume} {105}},\ \bibinfo
  {pages} {126011} (\bibinfo {year} {2022})},\ \Eprint
  {http://arxiv.org/abs/2112.11662} {arXiv:2112.11662 [hep-th]} \BibitemShut
  {NoStop}%
\bibitem [{\citenamefont {Jeong}\ \emph {et~al.}(2022)\citenamefont {Jeong},
  \citenamefont {Kim},\ and\ \citenamefont {Sun}}]{Jeong:2022luo}%
  \BibitemOpen
  \bibfield  {author} {\bibinfo {author} {\bibfnamefont {H.-S.}\ \bibnamefont
  {Jeong}}, \bibinfo {author} {\bibfnamefont {K.-Y.}\ \bibnamefont {Kim}}, \
  and\ \bibinfo {author} {\bibfnamefont {Y.-W.}\ \bibnamefont {Sun}},\ }\href
  {\doibase 10.1007/JHEP07(2022)065} {\bibfield  {journal} {\bibinfo  {journal}
  {JHEP}\ }\textbf {\bibinfo {volume} {07}},\ \bibinfo {pages} {065} (\bibinfo
  {year} {2022})},\ \Eprint {http://arxiv.org/abs/2203.02642} {arXiv:2203.02642
  [hep-th]} \BibitemShut {NoStop}%
\bibitem [{\citenamefont {Wang}\ and\ \citenamefont
  {Wang}(2022)}]{Wang:2022mcq}%
  \BibitemOpen
  \bibfield  {author} {\bibinfo {author} {\bibfnamefont {D.}~\bibnamefont
  {Wang}}\ and\ \bibinfo {author} {\bibfnamefont {Z.-Y.}\ \bibnamefont
  {Wang}},\ }\href {\doibase 10.1103/PhysRevLett.129.231603} {\bibfield
  {journal} {\bibinfo  {journal} {Phys. Rev. Lett.}\ }\textbf {\bibinfo
  {volume} {129}},\ \bibinfo {pages} {231603} (\bibinfo {year} {2022})},\
  \Eprint {http://arxiv.org/abs/2208.01047} {arXiv:2208.01047 [hep-th]}
  \BibitemShut {NoStop}%
\bibitem [{\citenamefont {Wang}\ and\ \citenamefont
  {Pan}(2023)}]{Wang:2022xoc}%
  \BibitemOpen
  \bibfield  {author} {\bibinfo {author} {\bibfnamefont {Y.-T.}\ \bibnamefont
  {Wang}}\ and\ \bibinfo {author} {\bibfnamefont {W.-B.}\ \bibnamefont {Pan}},\
  }\href {\doibase 10.1007/JHEP01(2023)174} {\bibfield  {journal} {\bibinfo
  {journal} {JHEP}\ }\textbf {\bibinfo {volume} {01}},\ \bibinfo {pages} {174}
  (\bibinfo {year} {2023})},\ \Eprint {http://arxiv.org/abs/2209.04296}
  {arXiv:2209.04296 [hep-th]} \BibitemShut {NoStop}%
\bibitem [{\citenamefont {Yuan}\ \emph {et~al.}(2023)\citenamefont {Yuan},
  \citenamefont {Ge}, \citenamefont {Kim}, \citenamefont {Ji},\ and\
  \citenamefont {Ahn}}]{Yuan:2023tft}%
  \BibitemOpen
  \bibfield  {author} {\bibinfo {author} {\bibfnamefont {H.}~\bibnamefont
  {Yuan}}, \bibinfo {author} {\bibfnamefont {X.-H.}\ \bibnamefont {Ge}},
  \bibinfo {author} {\bibfnamefont {K.-Y.}\ \bibnamefont {Kim}}, \bibinfo
  {author} {\bibfnamefont {C.-W.}\ \bibnamefont {Ji}}, \ and\ \bibinfo {author}
  {\bibfnamefont {Y.}~\bibnamefont {Ahn}},\ }\href {\doibase
  10.1007/JHEP08(2023)157} {\bibfield  {journal} {\bibinfo  {journal} {JHEP}\
  }\textbf {\bibinfo {volume} {08}},\ \bibinfo {pages} {157} (\bibinfo {year}
  {2023})},\ \Eprint {http://arxiv.org/abs/2303.04801} {arXiv:2303.04801
  [hep-th]} \BibitemShut {NoStop}%
\bibitem [{\citenamefont {Natsuume}\ and\ \citenamefont
  {Okamura}(2023)}]{Natsuume:2023lzy}%
  \BibitemOpen
  \bibfield  {author} {\bibinfo {author} {\bibfnamefont {M.}~\bibnamefont
  {Natsuume}}\ and\ \bibinfo {author} {\bibfnamefont {T.}~\bibnamefont
  {Okamura}},\ }\href {\doibase 10.1103/PhysRevD.108.046012} {\bibfield
  {journal} {\bibinfo  {journal} {Phys. Rev. D}\ }\textbf {\bibinfo {volume}
  {108}},\ \bibinfo {pages} {046012} (\bibinfo {year} {2023})},\ \Eprint
  {http://arxiv.org/abs/2306.03930} {arXiv:2306.03930 [hep-th]} \BibitemShut
  {NoStop}%
\bibitem [{\citenamefont {Baishya}\ \emph {et~al.}(2023)\citenamefont
  {Baishya}, \citenamefont {Chakrabarti},\ and\ \citenamefont
  {Maity}}]{Baishya:2023xbj}%
  \BibitemOpen
  \bibfield  {author} {\bibinfo {author} {\bibfnamefont {B.}~\bibnamefont
  {Baishya}}, \bibinfo {author} {\bibfnamefont {S.}~\bibnamefont
  {Chakrabarti}}, \ and\ \bibinfo {author} {\bibfnamefont {D.}~\bibnamefont
  {Maity}},\ }\href@noop {} {\  (\bibinfo {year} {2023})},\ \Eprint
  {http://arxiv.org/abs/2311.05314} {arXiv:2311.05314 [hep-th]} \BibitemShut
  {NoStop}%
\bibitem [{\citenamefont {Yadav}\ \emph {et~al.}(2024)\citenamefont {Yadav},
  \citenamefont {Kushwah},\ and\ \citenamefont {Misra}}]{Yadav:2023hyg}%
  \BibitemOpen
  \bibfield  {author} {\bibinfo {author} {\bibfnamefont {G.}~\bibnamefont
  {Yadav}}, \bibinfo {author} {\bibfnamefont {S.~S.}\ \bibnamefont {Kushwah}},
  \ and\ \bibinfo {author} {\bibfnamefont {A.}~\bibnamefont {Misra}},\ }\href
  {\doibase 10.1007/JHEP05(2024)015} {\bibfield  {journal} {\bibinfo  {journal}
  {JHEP}\ }\textbf {\bibinfo {volume} {05}},\ \bibinfo {pages} {015} (\bibinfo
  {year} {2024})},\ \Eprint {http://arxiv.org/abs/2311.09306} {arXiv:2311.09306
  [hep-th]} \BibitemShut {NoStop}%
\bibitem [{\citenamefont {Akutagawa}\ \emph {et~al.}(2019)\citenamefont
  {Akutagawa}, \citenamefont {Hashimoto}, \citenamefont {Murata},\ and\
  \citenamefont {Ota}}]{qcdchaos}%
  \BibitemOpen
  \bibfield  {author} {\bibinfo {author} {\bibfnamefont {T.}~\bibnamefont
  {Akutagawa}}, \bibinfo {author} {\bibfnamefont {K.}~\bibnamefont
  {Hashimoto}}, \bibinfo {author} {\bibfnamefont {K.}~\bibnamefont {Murata}}, \
  and\ \bibinfo {author} {\bibfnamefont {T.}~\bibnamefont {Ota}},\ }\href
  {\doibase 10.1103/PhysRevD.100.046009} {\bibfield  {journal} {\bibinfo
  {journal} {Phys. Rev. D}\ }\textbf {\bibinfo {volume} {100}},\ \bibinfo
  {pages} {046009} (\bibinfo {year} {2019})}\BibitemShut {NoStop}%
\bibitem [{\citenamefont {Akutagawa}\ \emph {et~al.}(2018)\citenamefont
  {Akutagawa}, \citenamefont {Hashimoto}, \citenamefont {Miyazaki},\ and\
  \citenamefont {Ota}}]{Akutagawa_2018}%
  \BibitemOpen
  \bibfield  {author} {\bibinfo {author} {\bibfnamefont {T.}~\bibnamefont
  {Akutagawa}}, \bibinfo {author} {\bibfnamefont {K.}~\bibnamefont
  {Hashimoto}}, \bibinfo {author} {\bibfnamefont {T.}~\bibnamefont {Miyazaki}},
  \ and\ \bibinfo {author} {\bibfnamefont {T.}~\bibnamefont {Ota}},\ }\href
  {\doibase 10.1093/ptep/pty055} {\bibfield  {journal} {\bibinfo  {journal}
  {Progress of Theoretical and Experimental Physics}\ }\textbf {\bibinfo
  {volume} {2018}} (\bibinfo {year} {2018}),\ 10.1093/ptep/pty055}\BibitemShut
  {NoStop}%
\bibitem [{\citenamefont {Shukla}\ \emph {et~al.}(2023)\citenamefont {Shukla},
  \citenamefont {Dudal},\ and\ \citenamefont {Mahapatra}}]{Shukla:2023pbp}%
  \BibitemOpen
  \bibfield  {author} {\bibinfo {author} {\bibfnamefont {B.}~\bibnamefont
  {Shukla}}, \bibinfo {author} {\bibfnamefont {D.}~\bibnamefont {Dudal}}, \
  and\ \bibinfo {author} {\bibfnamefont {S.}~\bibnamefont {Mahapatra}},\ }\href
  {\doibase 10.1007/JHEP06(2023)178} {\bibfield  {journal} {\bibinfo  {journal}
  {JHEP}\ }\textbf {\bibinfo {volume} {06}},\ \bibinfo {pages} {178} (\bibinfo
  {year} {2023})},\ \Eprint {http://arxiv.org/abs/2303.15716} {arXiv:2303.15716
  [hep-th]} \BibitemShut {NoStop}%
\bibitem [{\citenamefont {Karch}\ and\ \citenamefont
  {Katz}(2002)}]{Karch:2002sh}%
  \BibitemOpen
  \bibfield  {author} {\bibinfo {author} {\bibfnamefont {A.}~\bibnamefont
  {Karch}}\ and\ \bibinfo {author} {\bibfnamefont {E.}~\bibnamefont {Katz}},\
  }\href {\doibase 10.1088/1126-6708/2002/06/043} {\bibfield  {journal}
  {\bibinfo  {journal} {JHEP}\ }\textbf {\bibinfo {volume} {06}},\ \bibinfo
  {pages} {043} (\bibinfo {year} {2002})},\ \Eprint
  {http://arxiv.org/abs/hep-th/0205236} {arXiv:hep-th/0205236} \BibitemShut
  {NoStop}%
\bibitem [{\citenamefont {Maldacena}\ \emph {et~al.}(2016)\citenamefont
  {Maldacena}, \citenamefont {Shenker},\ and\ \citenamefont
  {Stanford}}]{Maldacena:2015waa}%
  \BibitemOpen
  \bibfield  {author} {\bibinfo {author} {\bibfnamefont {J.}~\bibnamefont
  {Maldacena}}, \bibinfo {author} {\bibfnamefont {S.~H.}\ \bibnamefont
  {Shenker}}, \ and\ \bibinfo {author} {\bibfnamefont {D.}~\bibnamefont
  {Stanford}},\ }\href {\doibase 10.1007/JHEP08(2016)106} {\bibfield  {journal}
  {\bibinfo  {journal} {JHEP}\ }\textbf {\bibinfo {volume} {08}},\ \bibinfo
  {pages} {106} (\bibinfo {year} {2016})},\ \Eprint
  {http://arxiv.org/abs/1503.01409} {arXiv:1503.01409 [hep-th]} \BibitemShut
  {NoStop}%
\bibitem [{\citenamefont {Kodama}\ and\ \citenamefont
  {Ishibashi}(2003)}]{Kodama:2003jz}%
  \BibitemOpen
  \bibfield  {author} {\bibinfo {author} {\bibfnamefont {H.}~\bibnamefont
  {Kodama}}\ and\ \bibinfo {author} {\bibfnamefont {A.}~\bibnamefont
  {Ishibashi}},\ }\href {\doibase 10.1143/PTP.110.701} {\bibfield  {journal}
  {\bibinfo  {journal} {Prog. Theor. Phys.}\ }\textbf {\bibinfo {volume}
  {110}},\ \bibinfo {pages} {701} (\bibinfo {year} {2003})},\ \Eprint
  {http://arxiv.org/abs/hep-th/0305147} {arXiv:hep-th/0305147} \BibitemShut
  {NoStop}%
\bibitem [{\citenamefont {Horowitz}\ and\ \citenamefont
  {Strominger}(1991)}]{Horowitz:1991cd}%
  \BibitemOpen
  \bibfield  {author} {\bibinfo {author} {\bibfnamefont {G.~T.}\ \bibnamefont
  {Horowitz}}\ and\ \bibinfo {author} {\bibfnamefont {A.}~\bibnamefont
  {Strominger}},\ }\href {\doibase 10.1016/0550-3213(91)90440-9} {\bibfield
  {journal} {\bibinfo  {journal} {Nucl. Phys. B}\ }\textbf {\bibinfo {volume}
  {360}},\ \bibinfo {pages} {197} (\bibinfo {year} {1991})}\BibitemShut
  {NoStop}%
\bibitem [{\citenamefont {Albash}\ \emph
  {et~al.}(2008{\natexlab{a}})\citenamefont {Albash}, \citenamefont {Filev},
  \citenamefont {Johnson},\ and\ \citenamefont {Kundu}}]{Albash:2007bk}%
  \BibitemOpen
  \bibfield  {author} {\bibinfo {author} {\bibfnamefont {T.}~\bibnamefont
  {Albash}}, \bibinfo {author} {\bibfnamefont {V.~G.}\ \bibnamefont {Filev}},
  \bibinfo {author} {\bibfnamefont {C.~V.}\ \bibnamefont {Johnson}}, \ and\
  \bibinfo {author} {\bibfnamefont {A.}~\bibnamefont {Kundu}},\ }\href
  {\doibase 10.1088/1126-6708/2008/07/080} {\bibfield  {journal} {\bibinfo
  {journal} {JHEP}\ }\textbf {\bibinfo {volume} {07}},\ \bibinfo {pages} {080}
  (\bibinfo {year} {2008}{\natexlab{a}})},\ \Eprint
  {http://arxiv.org/abs/0709.1547} {arXiv:0709.1547 [hep-th]} \BibitemShut
  {NoStop}%
\bibitem [{\citenamefont {Albash}\ \emph
  {et~al.}(2008{\natexlab{b}})\citenamefont {Albash}, \citenamefont {Filev},
  \citenamefont {Johnson},\ and\ \citenamefont {Kundu}}]{Albash:2007bq}%
  \BibitemOpen
  \bibfield  {author} {\bibinfo {author} {\bibfnamefont {T.}~\bibnamefont
  {Albash}}, \bibinfo {author} {\bibfnamefont {V.~G.}\ \bibnamefont {Filev}},
  \bibinfo {author} {\bibfnamefont {C.~V.}\ \bibnamefont {Johnson}}, \ and\
  \bibinfo {author} {\bibfnamefont {A.}~\bibnamefont {Kundu}},\ }\href
  {\doibase 10.1088/1126-6708/2008/08/092} {\bibfield  {journal} {\bibinfo
  {journal} {JHEP}\ }\textbf {\bibinfo {volume} {08}},\ \bibinfo {pages} {092}
  (\bibinfo {year} {2008}{\natexlab{b}})},\ \Eprint
  {http://arxiv.org/abs/0709.1554} {arXiv:0709.1554 [hep-th]} \BibitemShut
  {NoStop}%
\bibitem [{\citenamefont {Ceplak}\ \emph {et~al.}(2020)\citenamefont {Ceplak},
  \citenamefont {Ramdial},\ and\ \citenamefont {Vegh}}]{Ceplak:2019ymw}%
  \BibitemOpen
  \bibfield  {author} {\bibinfo {author} {\bibfnamefont {N.}~\bibnamefont
  {Ceplak}}, \bibinfo {author} {\bibfnamefont {K.}~\bibnamefont {Ramdial}}, \
  and\ \bibinfo {author} {\bibfnamefont {D.}~\bibnamefont {Vegh}},\ }\href
  {\doibase 10.1007/JHEP07(2020)203} {\bibfield  {journal} {\bibinfo  {journal}
  {JHEP}\ }\textbf {\bibinfo {volume} {07}},\ \bibinfo {pages} {203} (\bibinfo
  {year} {2020})},\ \Eprint {http://arxiv.org/abs/1910.02975} {arXiv:1910.02975
  [hep-th]} \BibitemShut {NoStop}%
\bibitem [{\citenamefont {Blake}\ \emph {et~al.}(2017)\citenamefont {Blake},
  \citenamefont {Davison},\ and\ \citenamefont {Sachdev}}]{Blake:2017qgd}%
  \BibitemOpen
  \bibfield  {author} {\bibinfo {author} {\bibfnamefont {M.}~\bibnamefont
  {Blake}}, \bibinfo {author} {\bibfnamefont {R.~A.}\ \bibnamefont {Davison}},
  \ and\ \bibinfo {author} {\bibfnamefont {S.}~\bibnamefont {Sachdev}},\ }\href
  {\doibase 10.1103/PhysRevD.96.106008} {\bibfield  {journal} {\bibinfo
  {journal} {Phys. Rev. D}\ }\textbf {\bibinfo {volume} {96}},\ \bibinfo
  {pages} {106008} (\bibinfo {year} {2017})},\ \Eprint
  {http://arxiv.org/abs/1705.07896} {arXiv:1705.07896 [hep-th]} \BibitemShut
  {NoStop}%
\bibitem [{\citenamefont {Policastro}\ \emph {et~al.}(2002)\citenamefont
  {Policastro}, \citenamefont {Son},\ and\ \citenamefont
  {Starinets}}]{Policastro:2002se}%
  \BibitemOpen
  \bibfield  {author} {\bibinfo {author} {\bibfnamefont {G.}~\bibnamefont
  {Policastro}}, \bibinfo {author} {\bibfnamefont {D.~T.}\ \bibnamefont {Son}},
  \ and\ \bibinfo {author} {\bibfnamefont {A.~O.}\ \bibnamefont {Starinets}},\
  }\href {\doibase 10.1088/1126-6708/2002/09/043} {\bibfield  {journal}
  {\bibinfo  {journal} {JHEP}\ }\textbf {\bibinfo {volume} {09}},\ \bibinfo
  {pages} {043} (\bibinfo {year} {2002})},\ \Eprint
  {http://arxiv.org/abs/hep-th/0205052} {arXiv:hep-th/0205052} \BibitemShut
  {NoStop}%
\bibitem [{\citenamefont {Banerjee}\ \emph {et~al.}(2019)\citenamefont
  {Banerjee}, \citenamefont {Kundu},\ and\ \citenamefont
  {Poojary}}]{Banerjee:2018kwy}%
  \BibitemOpen
  \bibfield  {author} {\bibinfo {author} {\bibfnamefont {A.}~\bibnamefont
  {Banerjee}}, \bibinfo {author} {\bibfnamefont {A.}~\bibnamefont {Kundu}}, \
  and\ \bibinfo {author} {\bibfnamefont {R.}~\bibnamefont {Poojary}},\ }\href
  {\doibase 10.1007/JHEP06(2019)076} {\bibfield  {journal} {\bibinfo  {journal}
  {JHEP}\ }\textbf {\bibinfo {volume} {06}},\ \bibinfo {pages} {076} (\bibinfo
  {year} {2019})},\ \Eprint {http://arxiv.org/abs/1811.04977} {arXiv:1811.04977
  [hep-th]} \BibitemShut {NoStop}%
\bibitem [{\citenamefont {Bigazzi}\ \emph {et~al.}(2013)\citenamefont
  {Bigazzi}, \citenamefont {Cotrone},\ and\ \citenamefont
  {Tarrio}}]{Bigazzi:2013jqa}%
  \BibitemOpen
  \bibfield  {author} {\bibinfo {author} {\bibfnamefont {F.}~\bibnamefont
  {Bigazzi}}, \bibinfo {author} {\bibfnamefont {A.~L.}\ \bibnamefont
  {Cotrone}}, \ and\ \bibinfo {author} {\bibfnamefont {J.}~\bibnamefont
  {Tarrio}},\ }\href {\doibase 10.1007/JHEP07(2013)074} {\bibfield  {journal}
  {\bibinfo  {journal} {JHEP}\ }\textbf {\bibinfo {volume} {07}},\ \bibinfo
  {pages} {074} (\bibinfo {year} {2013})},\ \Eprint
  {http://arxiv.org/abs/1304.4802} {arXiv:1304.4802 [hep-th]} \BibitemShut
  {NoStop}%
\end{thebibliography}%

\end{document}